\documentclass[aps,prb,reprint,floatfix,superscriptaddress]{revtex4-2}

\usepackage{amsmath,amsfonts,amssymb}
\usepackage{graphicx}
\usepackage{dcolumn}
\usepackage{bm}
\usepackage{braket}
\usepackage{xcolor}
\usepackage{tikz}
\usepackage[caption=false]{subfig}
\usepackage{mhchem}
\usepackage{microtype}
\usepackage{mathtools}

\usepackage{multirow}
\usepackage{makecell}
\usepackage{amsthm}
\usepackage{soul}
\usepackage{appendix}
\usepackage{hyperref}
\usepackage{subfig}
\usepackage{tabularx}
\usepackage{booktabs}

\newtheorem{lemma}{Lemma}

\newcommand{\bea}{\begin{eqnarray}}
\newcommand{\eea}{\end{eqnarray}}
\newcommand{\eq}[1]{Eq.~(\ref{#1})} %
\newcommand{\LL}{\mathfrak{L}}

\newcommand{\adjoint}[0]{\mathrm{ad}}
\newcommand{\norm}[1]{||#1||}

\begin{document}



\title{On the Feasibility of Exact Unitary Transformations for Many-body Hamiltonians}




\author{Praveen Jayakumar}
 \affiliation{Chemical Physics Theory Group, Department of Chemistry, University of Toronto, Toronto, Ontario M5S 3H6, Canada}
 \affiliation{Department of Physical and Environmental Sciences, University of Toronto Scarborough, Toronto, Ontario M1C 1A4, Canada }

\author{Tao Zeng}
 \affiliation{Department of Chemistry York University, Toronto,  Ontario M3J 1P3, Canada}

 \author{Artur F. Izmaylov}%
 \email{artur.izmaylov@utoronto.ca}
 \affiliation{Chemical Physics Theory Group, Department of Chemistry, University of Toronto, Toronto, Ontario M5S 3H6, Canada} \affiliation{Department of Physical and Environmental Sciences, University of Toronto Scarborough, Toronto, Ontario M1C 1A4, Canada }

\date{\today}
\begin{abstract}
Exact unitary transformations play a central role in the analysis and simulation of many-body quantum systems, yet the conditions under which they can be carried out exactly and efficiently remain incompletely understood. We show that exact transformations arise whenever the adjoint action of a unitary's generator defines a linear map within a finite-dimensional operator space. In this regime, there exists a finite-degree polynomial that annihilates the adjoint map, rendering the Baker-Campbell-Hausdorff (BCH) expansion finite. We identify the role of Lie algebras and their modules in producing finite BCH expansions in all known cases. This perspective brings together previously disparate examples of exact transformations under a single unifying principle and clarifies how algebraic relations between generators and transformed operators determine the polynomial degree of the transformation. We illustrate this framework for previously known cases of efficient unitary transformations including unitary coupled-cluster and Pauli product generators. Using this framework, we propose a new class of fermionic generators that can be used for efficient transformations. The result establishes sufficient algebraic conditions for when exact unitary transformations are possible and provides new strategies for reducing their computational cost in quantum simulation and constructing feasible unitary transformations.

\end{abstract}

\maketitle

\section{Introduction}

Unitary transformations of a Hamiltonian,
\begin{equation}
\hat H' = \hat U \hat H \hat U^\dagger,
\end{equation}
are central tools in quantum many-body theory and quantum algorithms. By systematically changing representation, they can simplify the structure of correlations and entanglement in the eigenstates of a Hamiltonian. In electronic structure theory, canonical transformations are used to approximately block-diagonalize Hamiltonians before truncating the Hilbert space, enabling downfolding approaches.\cite{White2002} In variational quantum eigensolver (VQE) algorithms,\cite{VQE} suitably chosen unitaries reduce circuit depth for ground- and excited-state preparation and enable efficient circuit cutting in distributed implementations.\cite{Schleich2023} In fault-tolerant settings, interaction-picture transformations are employed to reduce the cost of simulation algorithms.\cite{loaiza2023} Quantum-inspired classical techniques, such as iterative qubit coupled cluster (iQCC)\cite{ryabinkin2020} and Pauli propagation methods,\cite{rudolph2025} likewise rely on Hamiltonians transformed by simple unitaries in the Heisenberg picture.

Despite their ubiquity, it remains unclear under what conditions such transformations can be performed exactly and with polynomial computational cost. Evaluating the transformed Hamiltonian,
\begin{equation}
\hat H' = e^{i\theta \hat G} \hat H e^{-i\theta \hat G},
\end{equation}
typically involves an infinite Baker--Campbell--Hausdorff (BCH) expansion\cite{note:BCH},
\begin{equation}\label{eq:bch}
\hat H' = \sum_{j=0}^\infty \frac{\theta^j}{j!} \, \adjoint_{i\hat G}^j (\hat H),
\end{equation}
where the $j^{\mathrm{th}}$-order nested commutator of $i\hat G$ and $\hat H$ is
\begin{equation}
\adjoint_{i\hat G}^j(\hat H) = [i\hat G, \adjoint_{i\hat G}^{j-1}(\hat H)], \qquad \adjoint_{i\hat G}^0(\hat H) = \hat H.
\end{equation}
Direct computation of this series scales exponentially with the system size. Consequently, a central challenge is to identify situations where the series can be evaluated exactly using a few powers of the adjoint operation ($\adjoint_{i\hat G}(.)$).

A well-known simplification arises when the nested commutators close algebraically,
\begin{equation}\label{eq:finite}
\adjoint_{i\hat G}^n(\hat H) = \sum_{j=0}^{n-1} c_j\, \adjoint_{i\hat G}^j(\hat H),
\end{equation}
in which case the transformation requires only a finite number of terms. Many important examples of such algebraic closure are known, including fermionic orbital rotations,\cite{Helgaker:Book} Bogoliubov transformations for both fermions and bosons,\cite{Bogoliubov:Book} and Pauli-product unitaries.\cite{ryabinkin2020} Comprehensive treatments of these examples can be found in Ref.~\onlinecite{Wagner:Book}. Recent developments in quantum computing have extended this list to linear combinations of anti-commuting Pauli operators\cite{lang2021:ACP} and unitary coupled-cluster (UCC) generators.\cite{kottmann2021,izmaylov2021,evangelista2025} 

Although these examples demonstrate that algebraic closure can dramatically reduce computational cost, they are usually identified case by case. To date, no general algebraic condition has been established for when the BCH expansion can be expressed using a few terms due to the algebraic closure. Moreover, prior analyses have not clarified the mathematical structure underlying the operator relationships that make exact transformations possible.

In this work, we develop a unified algebraic framework that encompasses all known cases of finite BCH expansion and establishes the general condition for its termination. The central insight is to view the adjoint action $\adjoint_{i\hat G}(\cdot)$ as a linear map acting on a finite-dimensional operator space. The Cayley--Hamilton theorem then guarantees that this map satisfies a finite-degree characteristic polynomial, ensuring that any unitary transformation generated by such $\hat G$ is exactly representable by a polynomial of finite degree. This perspective unifies two key scenarios: when the generator $\hat G$ belongs to a finite-dimensional Lie algebra, and when it has a finite eigenspectrum. The framework further provides systematic guidance for exploiting additional algebraic properties of the transformed operator $\hat H$ to minimize the number of nested commutators required. Additionally, by enabling transformation under arbitrary generators of known small spectrum, the framework motivates construction of artificial generators with small spectrum and desired properties. As an example, we construct a class of generators with $\{\pm 1\}$ spectrum and are number preserving.

Finally, while our analysis naturally includes continuous families of transformations, we note that discrete unitary groups such as the Clifford group---which act as automorphisms of the Pauli group---are beyond the present scope. Optimization over such discrete structures involves qualitatively different considerations than the continuous settings addressed here.

\section{Theory}\label{sec:theory}

Before discussing algebraic closures for Hamiltonian transformations, it is useful to note a general property of any unitary transformation $\hat U$. When $\hat U$ is applied to an operator $\hat H$ that can be written as a polynomial of some operators $\{\hat O_i\}$,
\begin{equation}
    \hat H = \mathrm{poly}(\{\hat O_i\}),
\end{equation}
the transformed operator can be obtained by transforming each constituent operator individually:
\begin{equation}
    \hat U \hat H \hat U^\dagger = \mathrm{poly}(\{\hat U \hat O_i \hat U^\dagger\}).
\end{equation}
This observation shows that the transformation of a Hamiltonian reduces to understanding how the elementary operators $\hat O_i$ transform under conjugation.

\subsection{Algebraic closure from the Cayley--Hamilton theorem}

The algebraic closure condition in Eq.~(\ref{eq:finite}) follows from the existence of a finite-degree polynomial $p(\text{ad}_{i\hat G}) = 0$. The operation $\text{ad}_{i\hat G}(\hat O) = [i\hat G, \hat O]$ is a linear map, and when it is represented in a finite-dimensional linear space $V$, the Cayley--Hamilton theorem guarantees that this map is annihilated by its characteristic polynomial. Specifically, if $\mathbf{A}$ is the matrix representation of $\text{ad}_{i\hat G}$ in some basis of $V$, then the characteristic polynomial
\begin{equation}
    p_c(t) = \det(t\mathbf{1} - \mathbf{A})
\end{equation}
satisfies $p_c(\mathbf{A}) = 0$ and $\deg[p_c] = \dim(V)=d_V$. Consequently,
\begin{equation}
    (\text{ad}_{i\hat G})^{d_V} = p_{d_V-1}(\text{ad}_{i\hat G}),
\end{equation}
where $p_{d_V-1}$ is a polynomial of degree $d_V-1$. This relation directly yields the algebraic closure condition in Eq.~(\ref{eq:finite}) and ensures that the BCH series terminates.

To apply the Cayley--Hamilton theorem, one must represent $\text{ad}_{i\hat G}$ as a linear operator in a finite-dimensional space $V$. In what follows, we discuss two situations where such a finite representation exists. In both cases, the transformed operator $\hat O$ is represented by a vector $\vec{O}$ in $V$ as a linear combination of basis vectors $\{\vec{O}_i\}$, and $\text{ad}_{i\hat G}$ acts as a linear transformation that maps $\vec{O}$ to another vector $\vec{O}'$ in the same basis.

\subsection{Lie--algebraic representation}\label{sec:lie}

A particularly natural finite-dimensional representation of $\text{ad}_{i\hat G}$ arises when the generator $\hat G$ belongs to a finite Lie algebra $\mathfrak{L}$ with generating elements $\{\hat O_i\}_{i=1}^N$. In this case,
\begin{equation}
    \hat G = \sum_i c_i \hat O_i.
\end{equation}
We also assume that the Hamiltonian $\hat H$ can be expressed as a polynomial of $\mathfrak{L}$ elements, $\hat H = \mathrm{poly}(\{\hat O_i\})$. Although the associative product of operators $\hat O_i$ does not belong to $\mathfrak{L}$—only their non-associative product defined by commutators,
\begin{equation}
    [\hat O_i, \hat O_j] = \sum_k c_{ij}^{(k)} \hat O_k \in \mathfrak{L},
\end{equation}
does—the Hamiltonian is an element of the universal enveloping algebra (UEA) of $\mathfrak{L}$\cite{Barut:Book}. This poses no difficulty, since one only needs to transform each operator $\hat O_i \in \mathfrak{L}$ that enters $\hat H$.

The Lie algebra $\mathfrak{L}$ is itself a linear space, with elements represented as vectors $\vec{v} = \sum_i c_i \vec{O}_i$. The adjoint map $\text{ad}_{i\hat G}$ acts linearly on $\mathfrak{L}$,
\begin{equation}
    \text{ad}_{i\hat G}(\vec{O}_i) = \sum_k A_{ik} \vec{O}_k,
\end{equation}
where the constants $A_{ik}$ define the matrix representation of $\text{ad}_{i\hat G}$. The dimension of this representation is equal to the dimension of $\mathfrak{L}$, $\dim(\LL) = d_{\LL}$, providing an upper bound for the degree of the characteristic polynomial.

Further reduction of this degree is possible through known results from Lie algebra theory. For example, in an Abelian Lie algebra all elements commute, implying $\text{ad}_{i\hat G} = 0$, and the transformation is trivial, $\hat H' = \hat H$. More generally, Lie algebras fall into three categories: (1) semisimple, (2) nilpotent, and (3) solvable.\cite{Barut:Book,Gilmore:Book} In semisimple algebras, one can choose a basis of eigenvectors $\{\vec{O}_i'\}$ for $\text{ad}_{i\hat G}$,
\begin{equation}
    \text{ad}_{i\hat G}(\vec{O}_i') = o_i \vec{O}_i',
\end{equation}
where $o_i$ are either zero (commuting elements) or complex numbers (raising/lowering operators). In this representation, transforming $\hat H$ becomes straightforward if it is expressed as a polynomial of $\hat O_i'$. In nilpotent algebras, the degree of the characteristic polynomial cannot exceed the nilpotent class $n_c$ of the algebra, defined by the condition
\begin{equation}
    (\text{ad}_{i\hat G})^{n_c}(\vec{O}_i) = 0, \qquad \forall \, \hat G, \hat O_i \in \mathfrak{L}.
\end{equation}
For instance, in an Abelian algebra, $n_c=1$.

Without delving further into algebra classification, we can show that transforming $\hat H = \mathrm{poly}(\{\hat O_i\})$ by $\hat U = e^{i\theta \hat G}$ always yields
\begin{equation}
    \hat H' = \mathrm{poly}'(\{\hat O_i\}), \quad \deg[\mathrm{poly}(\cdot)] = \deg[\mathrm{poly}'(\cdot)].
\end{equation}
When $\hat G = \sum_i c_i \hat O_i$, conjugation by $\hat U$ defines an automorphism of $\mathfrak{L}$. Each basis element transforms linearly as
\begin{equation}\label{eqn:UaU}
    \hat U \hat O_\mu \hat U^\dagger = \sum_{\nu=1}^{d_{\LL}} [U]_{\nu\mu} \hat O_\nu,
\end{equation}
where $[U]_{\nu\mu}$ are the matrix elements of $\exp[\text{ad}_{i\theta\hat G}]$. To compute $U$, one can apply the BCH expansion from Eq.~(\ref{eq:bch}). Because the commutators close within the Lie algebra, the series can be resummed to a closed form. A faithful matrix representation $\rho: \mathfrak{L} \to \mathfrak{gl}(N)$, which preserves commutation relations and has a trivial kernel, allows this resummation explicitly:
\begin{align}
    \rho(e^{i\theta \hat G}\hat O_\mu e^{-i\theta\hat G}) &= \rho(\hat O_\mu) + i\theta[\rho(\hat G), \rho(\hat O_\mu)] + \dots \nonumber \\
    &= e^{i\theta \rho(\hat G)} \rho(\hat O_\mu) e^{-i\theta\rho(\hat G)}.
\end{align}
Consequently,
\begin{equation}\label{eq:matrix-transform}
    \rho(\hat U)\rho(\hat O_\mu)\rho(\hat U)^\dagger = \sum_{\nu=1}^{d_{\LL}} [U]_{\nu\mu} \rho(\hat O_\nu),
\end{equation}
where $\rho(\hat U) := \exp[i\theta\rho(\hat G)]$. Equating elements in Eq.~(\ref{eq:matrix-transform}) gives a system of $N^2$ linear equations of $\{U_{\nu\mu}\}_{\nu=1}^{d_{\mathfrak L}}$ for each $\mu =1,\dots, d_{\mathfrak L}$. The faithfulness of $\rho$ ensures that these systems have unique solutions.

\paragraph*{$\mathfrak{L}$-module extension.}
The preceding discussion assumes that the transformed operator belongs to the same Lie algebra as the generator. More generally, one can consider the transformation of an operator $\hat O$ that lies in a linear space $\mathcal{A}$ closed under the adjoint action of $\mathfrak{L}$ elements:
\begin{align}
    \hat O &= \sum_i c_i \hat A_i, \\
    [\hat O_j, \hat A_i] &= \sum_k a_{ji}^{(k)} \hat A_k.
\end{align}
The set $\{\hat A_i\}$ defines a basis of $\mathcal{A}$, and the coefficients $a_{ji}^{(k)}$ specify how $\mathfrak{L}$ acts on $\mathcal{A}$ under commutation. In this case, $\mathcal{A}$ is said to be an $\mathfrak{L}$-module. The adjoint action of $\hat G \in \mathfrak{L}$ on $\mathcal{A}$ can be written as
\begin{equation}\label{eqn:comm_mod}
    [\hat G, \hat A_i] = \sum_{j=1}^N [A]_{ij}\hat A_j,
\end{equation}
where $\mathbf{A}$ is the matrix representation of $\text{ad}_{i\hat G}$ on the module. Evaluating Eq.~(\ref{eqn:comm_mod}) for each $\hat A_i \in \mathcal{A}$ determines $\mathbf{A}$, and applying the BCH expansion yields
\begin{equation}\label{eqn:UaU_mod}
    e^{i\theta\hat G}\hat A_i e^{-i\theta\hat G} = \sum_{j=1}^N [e^{i\theta \mathbf{A}}]_{ji} \hat A_j.
\end{equation}
Thus, $[U]_{ji} = [e^{i\theta \mathbf{A}}]_{ji}$ in the notation of Eq.~(\ref{eqn:UaU}). The matrix exponential $e^{i\theta \mathbf{A}}$ defines the representation of the Lie group generated by $\mathfrak{L}$ on its module $\mathcal{A}$.

\subsection{Eigen\texorpdfstring{--}{--}subspace projector representation}

More generally, when $\hat G$ is an arbitrary Hermitian operator with real spectrum,
\begin{equation}\label{eqn:generator}
    \hat G = \sum_{i=1}^L g_i \hat P_i,
\end{equation}
where $\{g_i\}_{i=1}^L$ are eigenvalues and $\{\hat P_i\}$ are orthogonal projectors onto the corresponding eigenspaces, $\sum_i \hat P_i = \mathbf{1}$ and $\hat P_i \hat P_j = \delta_{ij}\hat P_i$, the adjoint action admits a simple block structure:
\begin{equation}\label{eqn:adjoint_power}
    \text{ad}_{i\hat G}(\hat O) \;=\; \sum_{j,k=1}^L i(g_j-g_k)\,\hat P_j \hat O \hat P_k.
\end{equation}
In this projector basis, the identity superoperator is
\begin{equation}
    \mathcal{I}(\,\cdot\,) \;=\; \sum_{i,j=1}^L \hat P_i (\,\cdot\,) \hat P_j,
\end{equation}
and the matrix representation of $\text{ad}_{i\hat G}$ has eigenvalues $\{\, i(g_j-g_k)\,\}_{j,k=1}^L$ on the blocks $\hat P_j(\cdot)\hat P_k$. The representation space has dimension $L^2$, but the degree of the characteristic (and minimal) polynomial of $\text{ad}_{i\hat G}$ is governed only by the number of distinct differences $g_j-g_k$. Thus, in the worst case (all differences distinct) the upper bound for the degree is $L(L-1)$, and degeneracies or vanishing blocks reduce this degree.

\paragraph*{$\mathfrak{L}$-module point of view.}
This construction can be viewed as an $\mathfrak{L}$-module, where $\mathfrak{L}$ is the Abelian Lie algebra generated by the projectors $\{\hat P_i\}$, $[\hat P_i,\hat P_j]=0$. Consider the linear space $\mathcal{O}$ with basis elements
\begin{equation}
    \hat O_{ij} := \hat P_i \hat O \hat P_j.
\end{equation}
Any operator $\hat O'=\sum_{i,j} c_{ij}\hat O_{ij}$ remains in $\mathcal{O}$ under the adjoint action, since $\text{ad}_{i\hat G}$ acts diagonally as in Eq.~(\ref{eqn:adjoint_power}).

\paragraph*{Unitary transformation via a finite polynomial in $\text{ad}_{i\hat G}$.}
We seek a finite polynomial representation of the conjugation,
\begin{equation}\label{eqn:bch_finite}
    e^{i\theta \hat G}\,\hat O\,e^{-i\theta \hat G} \;=\; \sum_{m=0}^{d} c_m(\theta)\, \big(\text{ad}_{i\hat G}\big)^m(\hat O),
\end{equation}
for suitable coefficients $\{c_m(\theta)\}_{m=0}^d$. Projecting Eq.~(\ref{eqn:bch_finite}) into the eigenspace blocks and using Eq.~(\ref{eqn:adjoint_power}) gives
\begin{align}
    \sum_{j,k} e^{i\theta(g_j-g_k)} \hat P_j \hat O \hat P_k
    \;=\; \sum_{j,k} \sum_{m=0}^{d} c_m(\theta)\, \big(i(g_j-g_k)\big)^m \hat P_j \hat O \hat P_k. \label{eqn:bch_expanded}
\end{align}
Let $\|\hat P_j \hat O \hat P_k\|=0$ denote a vanishing block. Because projectors onto distinct $(j,k)$ pairs are orthogonal, the nonvanishing blocks are linearly independent. Equating coefficients on those blocks yields, for each nonvanishing block,
\begin{equation}\label{eqn:bch_eqns}
    e^{i\theta \Delta_{jk}} \;=\; \sum_{m=0}^{d} c_m(\theta)\, (i(g_j-g_k))^m.
\end{equation}
We define the set of relevant differences supported by $\hat O$,
\begin{equation}\label{eqn:S}
    S \;:=\; \big\{\, g_j-g_k \;:\; \|\hat P_j \hat O \hat P_k\| \neq 0 \,\big\}.
\end{equation}
Collecting the equations in vector form with $\vec e \in \mathbb{C}^{|S|}$ and $\vec c \in \mathbb{C}^{d+1}$, where $\vec e = \{e^{i\theta \Delta}\}_{\Delta\in S}$ and $\vec c = \{c_m(\theta)\}_{m=0}^{d}$, we obtain
\begin{equation}\label{eqn:vandermonde}
    \vec e \;=\; W\, \vec c,
\end{equation}
with the generalized Vandermonde matrix $W \in \mathbb{C}^{|S|\times(d+1)}$ defined by \cite{gantmacher2025theory}
\begin{equation}
    [W]_{\Delta,m} \;=\; (i\Delta)^m, \qquad \Delta \in S,\; m=0,\dots,d. 
\end{equation}
Choosing $d=|S|-1$ makes $W$ square; since the rows correspond to distinct $\Delta\in S$, $W$ is invertible and the coefficients are uniquely determined by
\begin{equation}
    \vec c \;=\; W^{-1}\, \vec e.
\end{equation}
The invertibility of $W$ and uniqueness of $\vec c$ implies that $\{\big(\adjoint_{i\hat G}\big)^j(\hat O)\}_{j=0}^{|S| - 1}$ are linearly independent operators.
If $\hat O$ is Hermitian (and $\hat G$ is Hermitian), then $\text{ad}_{i\hat G}(\hat O)=i[\hat G,\hat O]$ is Hermitian, and so are $\big(\text{ad}_{i\hat G}\big)^m(\hat O)$ for all $m$. Therefore, the coefficients $\{c_m(\theta)\}$ are real (see Appendix~\ref{app:reality} for an explicit proof).

\paragraph*{Determining the nonzero blocks.}
The number of nested commutators and their coefficients depend on $S$ in Eq.~(\ref{eqn:S}). To determine which projected blocks $\hat P_j \hat O \hat P_k$ are nonvanishing, one can construct polynomials in $\text{ad}_{\hat G}$ that isolate the sum over blocks with a fixed eigenvalue difference $\Delta$. Let $\mathcal{D}$ denote the set of all differences $\{g_j-g_k\}$. The Lagrange interpolation polynomial
\begin{equation}\label{eqn:lagrange}
    \mathrm{Lag}_{\mathcal{D}}^{(\Delta)}(x) \;=\; \prod_{\substack{\delta \in \mathcal{D} \\ \delta \neq \Delta}} \frac{x-\delta}{\Delta-\delta}
\end{equation}
satisfies $\mathrm{Lag}_{\mathcal{D}}^{(\Delta)}(\Delta)=1$ and vanishes at every $\delta\in\mathcal{D}\setminus\{\Delta\}$. Using Eq.~(\ref{eqn:adjoint_power}), applying $\mathrm{Lag}_{\mathcal{D}}^{(\Delta)}\!\left(\text{ad}_{\hat G}\right)$ to $\hat O$ isolates the sum of blocks
\begin{equation}\label{eqn:Dl_blocks}
    \mathcal{B}_{\Delta} := \big\{\, \hat P_j \hat O \hat P_k \;|\; g_j-g_k=\Delta \,\big\}.
\end{equation}
If $\big\| \mathrm{Lag}_{\mathcal{D}}^{(\Delta)}\!\left(\text{ad}_{\hat G}\right)(\hat O) \big\| \neq 0$, then at least one block in $\mathcal{B}_{\Delta}$ is nonvanishing. Repeating this test over all $\Delta\in\mathcal{D}$ yields
\begin{equation}
    S \;=\; \big\{\, \Delta \in \mathcal{D} \;:\; \big\| \mathrm{Lag}_{\mathcal{D}}^{(\Delta)}\!\left(\text{ad}_{\hat G}\right)(\hat O) \big\| \neq 0 \,\big\}.
\end{equation}
We illustrate this test for Pauli product generators in Section \ref{sec:Pauli}.

\subsection{Block reduction due to additional algebraic structures}

While the Lagrange-based projector method described above provides a systematic way to detect vanishing blocks $\hat P_i \hat O \hat P_j$, it can become cumbersome in practice, particularly when there is flexibility in the choice of operators to be transformed. For instance, if the Hamiltonian can be expressed as a polynomial in some operators, $\hat H = \mathrm{poly}(\{\hat O_i\})$, one could equally well apply a linear transformation of the operator basis, $\hat O_i' = \sum_j c_{ij} \hat O_j$, obtaining $\hat H = \mathrm{poly}(\{\hat O_i'\})$. The transformed set $\{\hat O_i'\}$ may have fewer nonvanishing projected blocks $\hat P_k \hat O_i' \hat P_j$, thus making the transformation under $\hat U = e^{i\theta \hat G}$ more efficient. 

Exploiting additional algebraic relations such as commutation or anticommutation between $\hat G$ and $\hat O$ provides a direct route to identifying block sparsity without explicitly constructing the Lagrange projectors.

\subsubsection{Lie algebras}

If both $\hat G$ and $\hat O$ belong to the same Lie algebra $\mathfrak{L}$, then two different characteristic polynomials can be considered: (i) $p_{\mathfrak{L}}$, obtained from the Lie algebraic adjoint map $\text{ad}_{i\hat G}$ acting on $\mathfrak{L}$, and (ii) $p_{\Delta g}$, obtained from the eigen-decomposition of $\hat G$ assuming all projected blocks are nonvanishing. These two polynomials need not coincide: $p_{\mathfrak{L}}$ reflects the algebraic structure of $\LL$, while $p_{\Delta g}$ captures the spectral properties of $\hat G$ as expressed through its eigenvalues and projectors $\hat P_i$. A lower polynomial degree in the $\LL$ Lie-algebraic representation,
\[
    \deg(p_{\mathfrak{L}}) < \deg(p_{\Delta g}),
\]
implies that certain projected blocks $\hat P_i \hat O \hat P_j$ must vanish.

For semisimple Lie algebras, all generators can be chosen such that 
\begin{equation}
    [i\hat G, \hat O_i'] = o_i \hat O_i',
\end{equation}
where $o_i$ are either zero or purely imaginary numbers. This yields two distinct cases:
\begin{itemize}
    \item For $o_i = 0$, the operators commute with $\hat G$, and only diagonal blocks survive, $\hat P_j \hat O_i' \hat P_k = \delta_{jk}\,\hat P_j \hat O_i'$.
    \item For $o_i \neq 0$, the nonvanishing blocks satisfy $i(g_j - g_k) = o_i$. In this case, the degree of the characteristic polynomial of $\text{ad}_{i\hat G}$ acting on $\hat O_i'$ is 1.
\end{itemize}
For non-semisimple Lie algebras, not all generators can be diagonalized in this manner, and the analysis of block structure must proceed case by case.

\subsubsection{Anti-commutativity}

Anti-commutativity between $\hat G$ and $\hat O$ likewise leads to systematic block reductions. Using the spectral decomposition of $\hat G$ in Eq.~(\ref{eqn:generator}), the anti-commutator takes the form
\begin{equation}\label{eqn:anticomm}
    [\hat G, \hat O]_+ = \sum_{i,j} (g_i + g_j)\, \hat P_i \hat O \hat P_j.
\end{equation}
If $\hat G$ and $\hat O$ anticommute, i.e., $[\hat G,\hat O]_+ = 0$, then $\hat P_i \hat O \hat P_j = 0$ whenever $g_i + g_j \neq 0$. Consequently, the allowed eigenvalue differences $g_j - g_k$ over the nonvanishing projected blocks reduce to
\begin{equation}
    S = \{\, 2g \;|\; g,-g \in \mathrm{eig}(\hat G),\; \|\hat P_{g}\hat O\hat P_{-g}\| \neq 0 \,\}.
\end{equation}
Thus, in this case, $|S|$ is at most $L$, corresponding to the number of distinct eigenvalues of $\hat G$.

\section{Applications}

To illustrate the theoretical framework developed above, we now examine two representative classes of unitary transformations for which the adjoint action $\text{ad}_{i\hat G}$ operates within a finite-dimensional linear space. These examples demonstrate how the general conditions identified in Sec.~\ref{sec:theory} manifest in familiar physical contexts and lead to efficient, closed-form transformations.

\subsection{Lie algebras and their modules}

\subsubsection{Orbital rotations}\label{sec:u(N)}

This example involves orbital rotations generated by the $\mathfrak{u}(N)$ Lie algebra, which exemplify the case where both the generator and the transformed operators belong to the same finite Lie algebra or its modules. In this setting, the transformation can be represented exactly through matrix exponentiation, and the transformed Hamiltonian remains within the same algebraic structure. In practice, such rotations are routinely used to localize molecular orbitals (e.g., Foster–Boys, Edmiston–Ruedenberg, Pipek–Mezey), which typically increases sparsity and spatial locality in $\hat H$ by concentrating one- and two-electron integral weight and shrinking the support of fermionic strings.\cite{Boys1960,EdmistonRuedenberg1963,PipekMezey1989,Lehtola2014} The same unitary can also be absorbed classically into the integrals prior to VQE, rather than compiled as a circuit, thereby reducing circuit depth and often the effective Hamiltonian 1-norm (and associated measurement cost) in chemistry VQE workflows.\cite{Koridon2021,Gonthier2022,Okada2023} These are precisely the adjoint actions our framework handles in closed form via finite-dimensional representations.

The second-quantized electronic Hamiltonian is given by
\begin{equation}\label{eqn:H_elec}
    \hat H = \sum_{p,q=1}^N h_{pq}\,\hat a_p^\dagger \hat a_q 
    + \sum_{p,q,r,s=1}^N g_{pqrs}\,\hat a_p^\dagger \hat a_q \hat a_r^\dagger \hat a_s,
\end{equation}
where $N$ is the number of spin orbitals and 
$\mathcal{C} := \{\hat a_p^\dagger\}_{p=1}^{N}$, 
$\mathcal{A} := \{\hat a_p\}_{p=1}^{N}$ denote the fermionic creation and annihilation operators, respectively.
We consider unitary transformations of $\hat H$ generated by orbital rotations. 
Orbital rotations defined by a Hermitian matrix $M$ take the form
\begin{equation}\label{eqn:orbital_rotation}
    \hat V = \exp\!\left[i \sum_{p,q=1}^N [M]_{pq}\, \hat a_p^\dagger \hat a_q \right].
\end{equation}

The excitation operators 
$\mathcal{E} := \{\hat a_p^\dagger \hat a_q\}_{p,q=1}^{N}$ 
satisfy the commutation relation
\begin{equation}
    [\hat a_p^\dagger \hat a_q, \hat a_r^\dagger \hat a_s] 
    = \delta_{qr}\, \hat a_p^\dagger \hat a_s 
    - \delta_{ps}\, \hat a_r^\dagger \hat a_q,
\end{equation}
and therefore generate a $\mathfrak{u}(N)$ Lie algebra. 
The creation and annihilation operators transform under the adjoint action of this algebra according to
\begin{equation}\label{eqn:ferm_identity}
    [\hat a_p^\dagger \hat a_q, \hat a_r^\dagger] = \delta_{qr}\, \hat a_p^\dagger, 
    \qquad 
    [\hat a_p^\dagger \hat a_q, \hat a_r] = -\delta_{pr}\, \hat a_q,
\end{equation}
implying that $\mathcal{C}$ and $\mathcal{A}$ are $\mathfrak{u}(N)$-modules.
Since the Hamiltonian in Eq.~(\ref{eqn:H_elec}) is a noncommuting polynomial in both $\mathcal{E}$ and $\mathcal{C} \cup \mathcal{A}$, 
the transformed Hamiltonian $\hat H' = \hat V \hat H \hat V^\dagger$ 
can be obtained by either of the two methods described in Sec.~\ref{sec:lie}.
We outline both below.

\vspace{1ex}
\noindent
\textit{Matrix representation.}—
Consider the faithful (fundamental) representation of $\mathfrak{u}(N)$ defined by
\begin{equation}
    [\rho(\hat a_p^\dagger \hat a_q)]_{ab} = \delta_{ap}\, \delta_{bq},
\end{equation}
where $\delta_{ap}, \delta_{bq}$ are Kronecker delta functions. In this representation, $\rho(\hat V) = e^{iM}$. 
Solving Eq.~(\ref{eq:matrix-transform}) in this basis yields
\begin{equation}\label{eqn:ad_U(N)}
    \hat V \hat a_p^\dagger \hat a_q \hat V^\dagger 
    = \sum_{r,s=1}^N [U]_{(rs)(pq)}\, \hat a_r^\dagger \hat a_s,
\end{equation}
where the matrix elements of the adjoint action are
\begin{equation}
    [U]_{(rs)(pq)} = [e^{iM}]_{rp}\, [e^{-iM}]_{qs},
\end{equation}
and $(pq)$, $(rs)$ denote compound indices.
The transformed Hamiltonian $\hat H' = \hat V \hat H \hat V^\dagger$ 
retains the structure of Eq.~(\ref{eqn:H_elec}), with renormalized coefficients given by tensor contractions:
\begin{align}
    h_{pq}' &= \sum_{a,b=1}^N h_{ab}\, [U]_{(pq)(ab)}, \\
    g_{pqrs}' &= \sum_{a,b,c,d=1}^N g_{abcd}\, [U]_{(pq)(ab)} [U]_{(rs)(cd)}.
\end{align}

\vspace{1ex}
\noindent
\textit{Commutator-based derivation.}—
Using Eq.~(\ref{eqn:ferm_identity}), the commutators with the generator 
$\hat G = \sum_{p,q} [M]_{pq}\, \hat a_p^\dagger \hat a_q$ are
\begin{equation}
    [\hat G, \hat a_p^\dagger] = \sum_{q=1}^N [M]_{qp}\, \hat a_q^\dagger, 
    \qquad 
    [\hat G, \hat a_p] = -\sum_{q=1}^N [M]_{pq}\, \hat a_q.
\end{equation}
Using the BCH expansion, we find
\begin{equation}\label{eqn:U(N)_a}
    \hat V \hat a_p^\dagger \hat V^\dagger 
    = \sum_{q=1}^N [e^{iM}]_{qp}\, \hat a_q^\dagger, 
    \qquad 
    \hat V \hat a_p \hat V^\dagger 
    = \sum_{q=1}^N [e^{-iM}]_{pq}\, \hat a_q.
\end{equation}
The matrices $e^{\pm iM}$ constitute the defining representation of $U(N)$, and their products reproduce the adjoint transformation in Eq.~(\ref{eqn:ad_U(N)}), thus verifying consistency between the algebraic and spectral constructions.

\subsubsection{Heisenberg Lie algebra}

The Heisenberg algebra $\mathfrak{h}_N=\{\hat x_i,\hat p_i,\hat I\}_{i=1}^N$ provides a natural route for transforming bosonic Hamiltonians (polynomials in $\{\hat x_i,\hat p_i\}$) via displacement-type generators, in contrast to the orbital-rotation setting (Sec.~\ref{sec:u(N)}) for fermions. Its defining relations
\begin{equation}
    [\hat x_i,\hat p_j]=i\hbar\,\delta_{ij}\hat I,\qquad
    [\hat x_i,\hat x_j]=[\hat p_i,\hat p_j]=0
\end{equation}
imply that $\hat I$ is central and all double commutators vanish. Hence $\mathfrak{h}_N$ is nilpotent of class~2, and the adjoint superoperator $\text{ad}_{i\hat G}$ is nilpotent with index at most~2 on $\mathrm{span}\{\hat x_i,\hat p_i,\hat I\}$. Equivalently, the minimal (and characteristic) polynomial of $\text{ad}_{i\hat G}$ has degree $\le 2$, so the BCH expansion truncates exactly after the first commutator—independently of the number of modes.

For a linear generator
\begin{equation}
    \hat G=\sum_{i=1}^N\big(v_i\hat p_i - u_i\hat x_i\big)+z\hat I,\qquad u_i,v_i,z\in\mathbb{R},
\end{equation}
the adjoint actions are
\begin{equation}
    \text{ad}_{i\hat G}(\hat x_j)=\hbar v_j \hat I,\qquad
    \text{ad}_{i\hat G}(\hat p_j)=\hbar u_j \hat I,
\end{equation}
leading to the exact unitary transformations
\begin{align}
    e^{i\hat G}\hat x_j e^{-i\hat G}&=\hat x_j+\hbar v_j \hat I,\\
    e^{i\hat G}\hat p_j e^{-i\hat G}&=\hat p_j+\hbar u_j \hat I.
\end{align}
Thus, elements of the Heisenberg group act as displacements on $(\hat x_j,\hat p_j)$, and conjugation by any such linear $\hat G$ evaluates in closed form with only single commutators. Practically, this yields clear computational savings for bosonic problems: the adjoint action is captured within a finite-dimensional representation, and transformed polynomial Hamiltonians are obtained exactly without growing commutator depth.

\subsection{$\hat G$ eigensubspace decomposition}

\subsubsection{Single Pauli product}\label{sec:Pauli}

The generator for the unitary transformations 
used in a series of hybrid quantum-classical VQE\cite{QCC,qubitAdapt} and quantum inspired iQCC\cite{ryabinkin2020,iQCC-PT2} methods is the single Pauli product (string) $\hat G=\hat\sigma_1\otimes\cdots\otimes\hat\sigma_N$, with $\hat\sigma_j\in\{\hat e,\hat x,\hat y,\hat z\}$. $\hat G$ squares to the identity operator,
\begin{equation}\label{eq:involution}
    \hat G^2=\hat I,
\end{equation}
and is involutory. Consequently, $\hat G$ has spectrum $\{\pm 1\}$, and it is compactly represented as $\hat G = \hat P_{+} - \hat P_{-}$, where $\hat P_{\pm}$
are corresponding eigensubspace projectors. 

For a generic $\hat H$, the projected blocks need not vanish; thus $S=\{0,\pm 2\}$ (recall $S$ denotes the set of distinct eigenvalue differences on nonvanishing projected blocks). From Eq.~\eqref{eqn:bch_finite}, the transformed Hamiltonian can be written as
\begin{equation}\label{eqn:H_trans_0pm2}
    e^{i\theta \hat G}\hat H e^{-i\theta \hat G}
    = c_0\,\hat H + c_1\,[i\hat G,\hat H] + c_2\,[i\hat G,[i\hat G,\hat H]],
\end{equation}
with coefficients obtained from Eq.~\eqref{eqn:vandermonde} as
\begin{equation}\label{eqn:c_0pm2}
    (c_0,c_1,c_2)=\Bigl(1,\tfrac{1}{2}\sin 2\theta,\tfrac{1}{2}\sin^2\theta\Bigr).
\end{equation}

Using a suitable fermion–qubit mapping (e.g., Jordan–Wigner or Bravyi–Kitaev), the electronic Hamiltonian in Eq.~\eqref{eqn:H_elec} can be expressed as a sum of Pauli products,
\begin{equation}\label{eqn:H_pauli}
    \hat H=\sum_{\alpha=1}^{N_p} h_\alpha\,\hat\sigma^{(\alpha)},
\end{equation}
where $N_p$ is the number of Pauli strings and
$\hat\sigma^{(\alpha)}=\hat\sigma_1^{(\alpha)}\otimes\cdots\otimes\hat\sigma_N^{(\alpha)}$ with $\hat\sigma_j^{(\alpha)}\in\{\hat e,\hat x,\hat y,\hat z\}$. We transform each Pauli product in Eq.~\eqref{eqn:H_pauli} individually. Since any two Pauli products either commute or anticommute, transforming terms individually requires fewer commutators.

When $\hat G$ and $\hat\sigma^{(\alpha)}$ commute, $S=\{0\}$, so $\hat\sigma^{(\alpha)\prime}=\hat\sigma^{(\alpha)}$. When they anticommute, we have
\[
    \hat P_+\,\hat\sigma^{(\alpha)}\,\hat P_+=\hat P_-\,\hat\sigma^{(\alpha)}\,\hat P_-=0,
\]
i.e., $\hat\sigma^{(\alpha)}$ is block off-diagonal in the eigenbasis of $\hat G$, and $S=\{\pm 2\}$, resulting in
\begin{equation}\label{eqn:H_trans_pm2}
    \hat\sigma^{(\alpha)\prime}
    = c_0\,\hat\sigma^{(\alpha)} + c_1\,[i\hat G,\hat\sigma^{(\alpha)}],
\end{equation}
with coefficients from Eq.~\eqref{eqn:vandermonde} given by
\begin{equation}\label{eqn:c_pm2}
    (c_0,c_1)=\Bigl(\cos 2\theta,\tfrac{1}{2}\sin 2\theta\Bigr).
\end{equation}
The above analysis can be carried out with Lagrange polynomials of $\adjoint_{\hat G}$. Since the eigenvalues of $\hat G$ are $\pm 1$, $\mathcal D = \{-2, 0, 2\}$. Then
\begin{align}
    \mathrm{Lag}_{\mathcal D}^{(-2)}(\adjoint_{\hat G})(\hat \sigma^{(\alpha)}) =& \tfrac{1}{8}\adjoint_{\hat G}^2(\hat \sigma^{(\alpha)}) - \tfrac{1}{4}\adjoint_{\hat G}(\hat \sigma^{(\alpha)}),\label{eq:poly_-2}\\
    \mathrm{Lag}_{\mathcal D}^{(0)}(\adjoint_{\hat G})(\hat \sigma^{(\alpha)}) =& -\tfrac{1}{4}\adjoint_{\hat G}^2(\hat \sigma^{(\alpha)}) + \hat \sigma^{(\alpha)},\label{eq:poly_0}\\
    \mathrm{Lag}_{\mathcal D}^{(2)}(\adjoint_{\hat G})(\hat \sigma^{(\alpha)}) =& \tfrac{1}{8}\adjoint_{\hat G}^2(\hat \sigma^{(\alpha)}) + \tfrac{1}{4}\adjoint_{\hat G}(\hat \sigma^{(\alpha)}).\label{eq:poly_2} 
\end{align}
When $\hat \sigma^{(\alpha)}$ commutes with $\hat G$ ($\adjoint_{\hat G}(\hat \sigma^{(\alpha)}) = 0$), Eqs. \eqref{eq:poly_-2}, \eqref{eq:poly_2} vanish while Eq. \eqref{eq:poly_0} does not, leading to $S = \{0\}$. When $\hat \sigma^{(\alpha)}$ anti-commutes with $\hat G$, we have $[\hat G, \hat \sigma^{(\alpha)}] = 2\hat G\hat \sigma^{(\alpha)}$ and Eqs. \eqref{eq:poly_-2}, \eqref{eq:poly_2} does not vanish while Eq. \eqref{eq:poly_0} vanishes due to Eq. \eqref{eq:involution}. This yields $S = \{-2, 2\}$.

\subsubsection{Mutually Anticommuting Pauli Products}

A generalization of a single Pauli-product generator is the sum of mutually anticommuting Pauli products $\{\hat A_j\}$\,\cite{lang2021:ACP}:
\begin{equation}
    \hat G=\sum_{j} d_j\,\hat A_j,\qquad d_j\in\mathbb{R},
\end{equation}
with
\begin{equation}
    \{\hat A_j,\hat A_k\}=0\ (j\neq k).
\end{equation}
It follows that
\begin{equation}
    \hat G^2=\Bigl(\sum_{j} d_j^2\Bigr)\hat I,
\end{equation}
so $\hat G$ is involutory when $\sum_{j} d_j^2=1$, and its spectrum is $\{\pm1\}$. The eigensubspace-projector representation then coincides with the single Pauli-product case, $\hat G=\hat P_{+}-\hat P_{-}$,
and Eq.~\eqref{eqn:H_trans_0pm2} holds exactly.

However, obtaining the single-commutator reduction of Eq.~\eqref{eqn:H_trans_pm2} is more restrictive: for a Pauli product $\hat\sigma^{(\alpha)}$ in $\hat H$ to commute (respectively, anticommute) with $\hat G$, it must commute (respectively, anticommute) with every $\hat A_j$. Consequently, not every Pauli product in $\hat H$ can be transformed with a single commutator with $\hat G$; in general, the quadratic term in Eq.~\eqref{eqn:H_trans_0pm2} remains necessary.

\subsubsection{Unitary Coupled Cluster Generators}\label{sec:ucc}

A unitary coupled-cluster (UCC) generator is the sum of a fermionic excitation and its corresponding de-excitation. It has been widely used in VQE workflows\cite{VQE,kottmann2021,Bauman_2019} and quantum-inspired algorithms.\cite{UCC-QI,UCC-QI2}
To discuss algebraic properties of UCC generators, we introduce the notation
\begin{equation}\label{eqn:ferm_str_notation}
    \hat a_A := \prod_{j\in A} \hat a_j, \qquad \hat a_A^\dagger := (\hat a_A)^\dagger,
\end{equation}
for products of fermionic annihilation and creation operators over a subset \(A\) of spin-orbital indices (taken in increasing order to fix phases). Then a UCC generator has the form
\begin{equation}\label{eqn:ucc_G}
    \hat G = (-i)\bigl(\hat T_G - \hat T_G^\dagger\bigr),
\end{equation}
where \(\hat T_G = \hat a_A^\dagger \hat a_B\) is an excitation with \(|A|=|B|\) and \(A\cap B=\varnothing\).  \(\hat G\) has spectrum \(\{0,\pm 1\}\).\cite{kottmann2021} Assuming no projected blocks vanish for a generic \(\hat H\), the set of distinct eigenvalue differences on nonvanishing blocks is \(S=\{0,\pm 1,\pm 2\}\). A polynomial of degree \(|S|-1=4\) (i.e., up to the fourth nested commutator) suffices:
\begin{equation}
    \hat H' = \sum_{j=0}^4 c_j\,\text{ad}_{i\hat G}^{\,j}(\hat H)
\end{equation}
with coefficients from Eq.~\eqref{eqn:vandermonde}:
\begin{align}
    c_0 =&\, 1,\\
    c_1 =&\, \tfrac{4}{3}\sin\theta - \tfrac{1}{6}\sin 2\theta,\\ 
    c_2 =&\, \tfrac{5}{4} - \tfrac{4}{3}\cos\theta + \tfrac{1}{12}\cos 2\theta,\\
    c_3 =&\, \tfrac{1}{3}\sin\theta - \tfrac{1}{6}\sin 2\theta,\\
    c_4 =&\, \tfrac{1}{4} - \tfrac{1}{3}\cos\theta + \tfrac{1}{12}\cos 2\theta.
\end{align}

Motivated by the structure of the electronic Hamiltonian and the UCC generator, we consider fragments of the Hamiltonian of the form
\begin{equation}\label{eq:ucc_fragment}
    \hat H_\alpha = \hat T_\alpha + \hat T_\alpha^\dagger,
\end{equation}
where \(\hat T_\alpha = \hat a_C^\dagger \hat a_D\) is a single fermionic string. Let \(\hat P_\pm, \hat P_0\) denote the projectors onto the eigenspaces \(\mathcal V_\pm, \mathcal V_0\), corresponding to eigenvalues \(\pm 1, 0\) of \(\hat G\), respectively. In Appendix~\ref{app:ucc_blocks}, we show that for any \(\hat T_\alpha\), either the set \(\{\hat P_\pm \hat H_\alpha \hat P_\pm,\;\hat P_\pm \hat H_\alpha \hat P_\mp\}\) or the set \(\{\hat P_\pm \hat H_\alpha \hat P_0,\;\hat P_0 \hat H_\alpha \hat P_\pm\}\) is potentially nonvanishing, but not both simultaneously. 

When the blocks \(\{\hat P_\pm \hat H_\alpha \hat P_\pm,\;\hat P_\pm \hat H_\alpha \hat P_\mp,\;\hat P_0 \hat H_\alpha \hat P_0\}\) are nonvanishing, \(S=\{0,\pm 2\}\) and the transformed fragment is given by Eq.~\eqref{eqn:H_trans_0pm2} with coefficients in Eq.~\eqref{eqn:c_0pm2}. When the blocks \(\{\hat P_\pm \hat H_\alpha \hat P_0,\;\hat P_0 \hat H_\alpha \hat P_\mp,\;\hat P_0 \hat H_\alpha \hat P_0\}\) are nonvanishing, \(S=\{0,\pm 1\}\) and the transformed fragment is again given by Eq.~\eqref{eqn:H_trans_0pm2}, but with coefficients from Eq.~\eqref{eqn:vandermonde}:
\begin{equation}
    (c_0, c_1, c_2) = \bigl(1,\, \sin\theta,\, 1 - \cos\theta\bigr).
\end{equation}
This reduction in projected blocks thus far remain true for the transformation of a single (non Hermitian) Fermionic excitation operator $\hat O_\alpha = \hat T_\alpha$ and does not require number preservation ($|A| = |B|$). Hence our analysis agrees with findings of Ref.~\cite{evangelista2025}. 

In addition, for the Hermitian fragments in Eq.~\eqref{eq:ucc_fragment}, we identify two further cases in which a single commutator suffices. In one case, the blocks \(\hat P_\pm \hat H_\alpha \hat P_\pm\), \(\hat P_\pm \hat H_\alpha \hat P_\mp\), and \(\hat P_0 \hat H_\alpha \hat P_0\) vanish, leading to \(S=\{\pm 1\}\). The transformed fragment then has the form of Eq.~\eqref{eqn:H_trans_pm2}, with coefficients from Eq.~\eqref{eqn:vandermonde}:
\begin{equation}
    (c_0, c_1) = \bigl(\cos\theta,\, \sin\theta\bigr).
\end{equation}
In the other case, the blocks \(\hat P_\pm \hat H_\alpha \hat P_\pm\), \(\hat P_0 \hat H_\alpha \hat P_0\), \(\hat P_\pm \hat H_\alpha \hat P_0\), and \(\hat P_0 \hat H_\alpha \hat P_\pm\) vanish. This yields \(S=\{\pm 2\}\) and a transformed fragment given by Eq.~\eqref{eqn:H_trans_pm2} with coefficients from Eq.~\eqref{eqn:c_pm2}. In both of these cases, \(|S|=2\) and only a single nested commutator is required to represent \(\hat H'_\alpha\).

\begin{figure}
    \centering
    \includegraphics[width=\columnwidth]{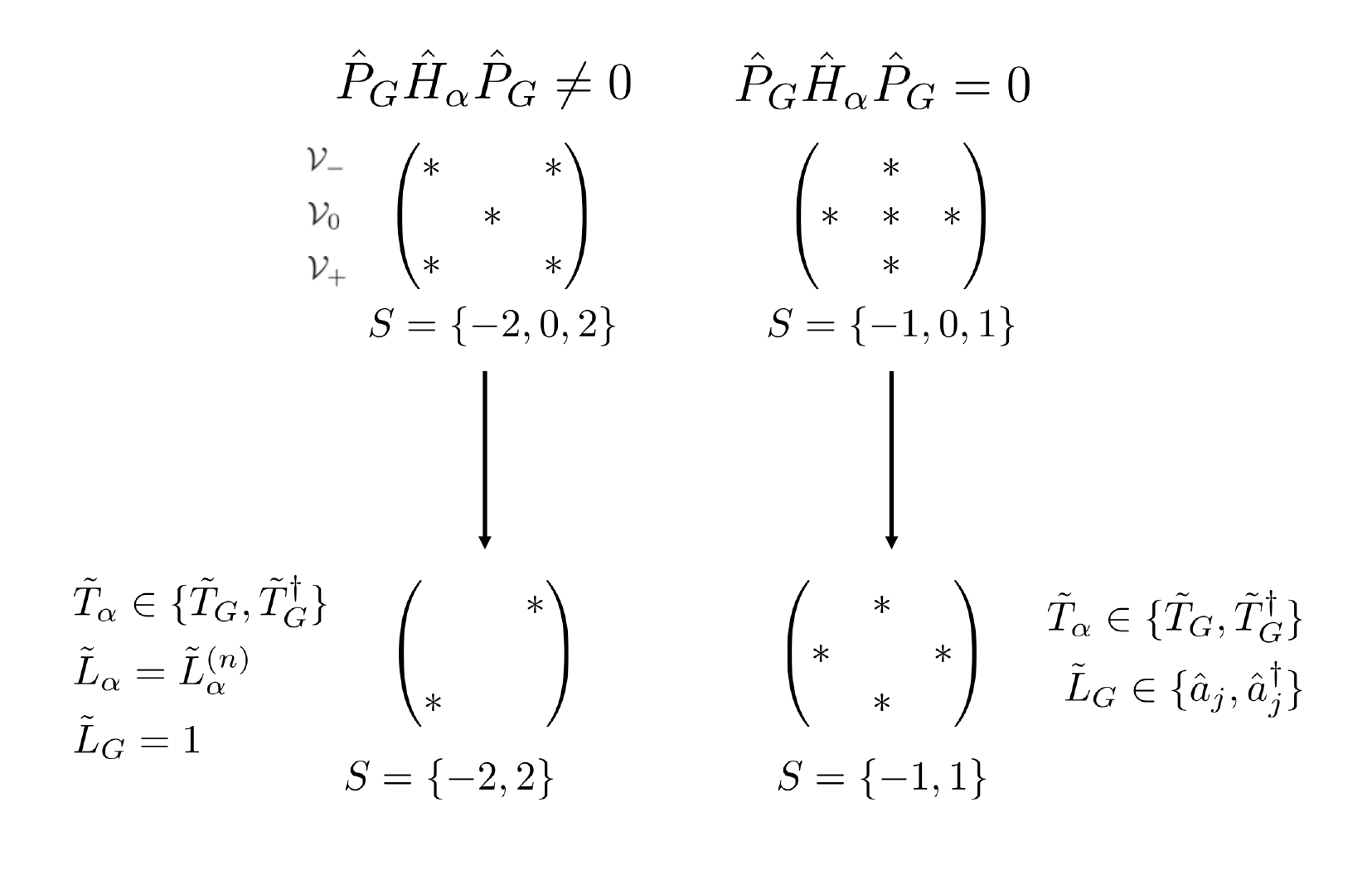}
    \caption{Two patterns of nonvanishing projected blocks (depicted by \(*\)) of \(\hat H_\alpha=\hat T_\alpha+\hat T_\alpha^\dagger\) in the eigenspaces \(\{\mathcal V_-,\mathcal V_0,\mathcal V_+\}\) of \(\hat G=(-i)(\hat T_G-\hat T_G^\dagger)\), depending on whether \(\hat P_G \hat H_\alpha \hat P_G\) vanishes, where \(\hat P_G := \hat P_+ + \hat P_-\). When the operators \(\hat T_G := \tilde T_G \tilde L_G\) and \(\hat T_\alpha := \tilde T_\alpha \tilde L_\alpha\) take the forms specified, additional projected blocks vanish. A superscript \((n)\) denotes a product of occupation operators, \(\{\hat a_p^\dagger \hat a_p\}_{p=1}^N\).}
    \label{fig:p_blocks}
\end{figure}

The set of vanishing projected blocks depends on the spin-orbital indices present in \(\hat T_\alpha\) and those shared with \(\hat T_G\). To list and analyze all possibilities, we introduce the notation
\begin{align}
    \hat T_G \;=&\; \tilde T_G\, \tilde L_G\label{eqn:tg_decomp}\\
    \hat T_\alpha \;=&\; \tilde T_\alpha\, \tilde L_\alpha\label{eqn:ta_decomp}
\end{align}
where \(\tilde T_G, \tilde T_\alpha\) act on the shared spin-orbital indices \((A\cup B)\cap(C\cup D)\), while \(\tilde L_G\) and \(\tilde L_\alpha\) act on the disjoint sets \((A\cup B)\setminus(C\cup D)\) and \((C\cup D)\setminus(A\cup B)\), respectively. Conditions on \(\{\tilde T_G, \tilde L_G, \tilde T_\alpha, \tilde L_\alpha\}\) that produce the various nonvanishing block structures described above are derived in Appendix~\ref{app:ucc_blocks}. The cases with examples are summarized in Table~\ref{tab:examples}, and Fig.~\ref{fig:p_blocks} illustrates their projected-block patterns.

\begin{table*}[]
    \centering
    \renewcommand{\arraystretch}{1.4}
    \begin{tabular}{|c|c|c|>{\centering\arraybackslash}p{2cm}>{\centering\arraybackslash}p{2cm}|}
    \hline
         Constraints & Vanishing blocks & $S$ & $\hat T_G$ & $\hat T_\alpha$\\\hline
         $\tilde T_\alpha \in \{\tilde T_G, \tilde T_G^\dagger\}$, $\tilde L_G = \tilde L_G^{(n)}$ & \multirow{2}{*}{$\hat P_\pm \hat H_\alpha \hat P_0$, $\hat P_0 \hat H_\alpha \hat P_\pm$} & \multirow{2}{*}{$\{-2, 0, 2\}$} & $\hat a_2^\dagger \hat a_1^\dagger \hat a_2 \hat a_0$ & $\hat a_3^\dagger \hat a_1^\dagger \hat a_4 \hat a_0$\\\cline{1-1}\cline{4-5}
         $\tilde T_\alpha \in \{\tilde T_G \tilde T_G^\dagger, \tilde T_G^\dagger \tilde T_G\}$ & & & $\hat a_3^\dagger \hat a_2^\dagger \hat a_1 \hat a_0$ & $\hat a_1^\dagger \hat a_1$\\\hline
        $\tilde T_\alpha \in \{\tilde T_G, \tilde T_G^\dagger\}$, $\tilde L_G = 1$, $\tilde L_\alpha = \tilde L_\alpha^{(n)}$ & $\hat P_\pm \hat H_\alpha \hat P_\pm$, $\hat P_0 \hat H_\alpha \hat P_0$, $\hat P_\pm \hat H_\alpha \hat P_0$, $\hat P_0 \hat H_\alpha \hat P_\pm$ & $\{-2, 2\}$ & $\hat a_3^\dagger \hat a_2^\dagger \hat a_1 \hat a_0$ & $\hat a_3^\dagger \hat a_2^\dagger \hat a_1 \hat a_0$\\\hline
        $\tilde T_\alpha \in \{\tilde T_G, \tilde T_G^\dagger\}$, $\tilde L_G \in \{\hat a_j, \hat a_j^\dagger\}$, $j\notin C\cup D$ & $\hat P_\pm \hat H_\alpha \hat P_\pm$, $\hat P_\pm \hat H_\alpha \hat P_\mp$, $\hat P_0 \hat H_\alpha \hat P_0$ & $\{-1, 1\}$ & $\hat a_3^\dagger \hat a_2^\dagger \hat a_1 \hat a_0$ & $\hat a_4^\dagger \hat a_2^\dagger \hat a_1 \hat a_0$\\\hline
        $\hat P_G \hat H_\alpha \hat P_G = 0$ and not above-listed cases & $\hat P_\pm \hat H_\alpha \hat P_\pm$, $\hat P_\pm \hat H_\alpha \hat P_\mp$ & $\{-1, 0, 1\}$ & $\hat a_3^\dagger \hat a_2^\dagger \hat a_1 \hat a_0$ & $\hat a_3^\dagger \hat a_2^\dagger \hat a_2 \hat a_0$\\\hline
    \end{tabular}
    \renewcommand{\arraystretch}{1.0}
    \caption{Examples of \(\{\hat T_G, \hat T_\alpha\}\) pairs resulting in different vanishing block structures and eigenvalue-difference sets \(S\), with the notation introduced in Eqs.~\eqref{eqn:tg_decomp} and \eqref{eqn:ta_decomp}. A superscript \((n)\) denotes a product of occupation operators, \(\{\hat a_p^\dagger \hat a_p\}_{p=1}^N\).}
    \label{tab:examples}
\end{table*}

\section{New many-body transformations}

The proposed framework not only encompasses all the known exact transformations but also enables us to create a new class of exact transformations. This new class is motivated by the desire to combine strengths of Pauli product based generators (lowest number of eigenvalues) with the electron number preserving character of the fermionic generators. Indeed, neither a single Pauli product nor a sum of mutually anticommuting Pauli products preserves the fermionic 
number symmetry of the transformed Hamiltonian. We introduce fermionic, electron number preserving, 
generators
\begin{equation}\label{eq:VpV}
    \hat G = \hat V \hat p\hat V^\dagger,
\end{equation}
where $\hat V$ is an orbital rotation [\eq{eqn:orbital_rotation}] and 
\begin{equation}
    \hat p = c_0 + \sum_p c_p\hat n_p + \sum_{pq}c_{pq}\hat n_p\hat n_q + \cdots
\end{equation}
is a polynomial in the occupation operators $\hat n_p=\hat a_p^\dagger \hat a_p$ with only a few distinct eigenvalues. 
This form is motivated by the ease of constructing polynomials with a fixed number of eigenvalues from the simplest hermitian fermionic projectors, $\{\hat n_p\}$. The orbital rotations $\hat V$ do not alter either the eigenvalues determined by $\hat p$ or fermionic rank of $\hat p$. The form of $\hat p$ and $\hat V$ can be chosen to satisfy other desired properties, such as preservation of spin symmetries. Interestingly, a UCC generator cannot be expressed in the form of Eq.~\eqref{eq:VpV}.\footnote{A UCC generator is a linear combination of commuting Pauli products (under any fermion-qubit mapping). Hence, it can be diagonalized by a Clifford unitary rather than by an orbital rotation.}

There are several approaches for constructing operators $\hat p$ with the target eigenvalues $g_i \in \Lambda$. A simple approach is to construct orthogonal projectors $\hat P_i$ as products of $\{\hat n_p, 1 - \hat n_p\}$. The generator is then constructed as a sum of projectors with the eigenvalues as coefficients. For example, to construct a fermionic rank $2$ generator with four eigenvalues $\{g_{1-4}\}$, the projectors are constructed as $\{\hat n_1\hat n_2, \hat n_1(1 - \hat n_2), (1 - \hat n_1)\hat n_2, (1 - \hat n_1)(1 - \hat n_2)\}$ and the generator takes the form
\begin{align}
    \hat p =& g_1\hat n_1\hat n_2 + g_2\hat n_1(1 - \hat n_2)\nonumber \\+& g_3(1 - \hat n_1)\hat n_2 + g_4(1 - \hat n_1)(1 - \hat n_2).
\end{align}
When the number of distinct spin-orbital indices is greater than the desired fermionic rank of the generator, some of the constructed projectors are no longer orthogonal. In such cases, this method cannot be used.

Alternatively, one can solve for coefficients $\{c_{p\cdots q}\}$ in $\hat p$ that satisfy the characteristic equation $p_\Lambda(\hat p) = \hat 0$. Here $p_\Lambda$ is the characteristic polynomial of the generator, and is defined as
\begin{equation}\label{eq:gen_characteristic}
    p_{\Lambda}(x)= \prod_{i} (x - g_i).
\end{equation}
The constraint $p_\Lambda(\hat p) = \hat 0$ yields a system of nonlinear equations in the coefficients, which can be solved analytically. Since non-trivial $\hat p$ can satisfy factors $p_{\Lambda'}(\hat p) =\hat 0$ for $\Lambda' \subset \Lambda, |\Lambda'| \geq 2$, such solutions do not exhibit the spectrum $\Lambda$ and must be discarded.

\subsection{Fermionic reflections}
To construct up to two-electron generators (fermionic rank $2$) with eigenvalues of $\pm 1$, the only operators $\hat p$ (up to index relabeling and a global phase of $\pm 1$) are
\begin{align}
 \hat p_1 &= 1 - 2\hat n_p \nonumber \\
 \hat p_2 &= 1 - 2\hat n_p \hat n_q \nonumber \\
 \hat p_3 &= 1 - 2\hat n_p + 2 \hat n_p \hat n_q \nonumber \\
 \hat p_4 &= 1 - 2\hat n_p - 2\hat n_q + 2\hat n_p\hat n_q \nonumber \\
  \hat p_5 &= 1 - 2\hat n_p - 2\hat n_q + 4 \hat n_p\hat n_q \nonumber \\
 \hat p_6  &= 1 - 2\hat n_p - 2\hat n_q\hat n_r + 2\hat n_p\hat n_r \nonumber \\
 \hat p_7  &= 1 - 2\hat n_p - 2\hat n_q\hat n_r + 2\hat n_p\hat n_r + 2\hat n_p\hat n_q \nonumber \\
 \hat p_8 &= 1 - 2\hat n_p - 2\hat n_q + 2\hat n_p\hat n_r + 2\hat n_p\hat n_q \nonumber \\
 \hat p_9 &= 1 - 2\hat n_p -2 \hat n_q -2\hat n_r + 2\hat n_q\hat n_r \nonumber \\
 &+ 2\hat n_p\hat n_r + 2\hat n_p\hat n_q \nonumber \\
 \hat p_{10} &= 1 - 2\hat n_p -2\hat n_q -2\hat n_r\hat n_s +2\hat n_p\hat n_r \nonumber \\ &+2 \hat n_q\hat n_s + 2\hat n_p\hat n_q.\label{eq:refl_poly_all}
\end{align}
These solutions are obtained by imposing $(c_0 + \sum_pc_p\hat n_p + \sum_{pq}c_{pq}\hat n_p\hat n_q)^2 - \hat 1=0$ and solving exhaustively for nontrivial values of $c_0, \{c_p\}, \{c_{pq}\}$. We find that $c_0 = 1$, $c_p \in \{0, -2\}$ and $c_{pq}\in \{0, \pm 2, 4\}$, and that there are no solutions involving $5$ distinct spin-orbital indices. Thus, no solutions exist with more than four distinct spin-orbital indices. The $10$ polynomials form disjoint classes of generators because orbital rotations cannot map one of these polynomials to another (see Appendix \ref{app:ferm_refl} for a proof). Since the constructed generators are reflective ($\hat G^2 = \hat 1$), we refer to these generators as fermionic reflections.

By selecting the polynomial $\hat p$ and matrix $M$ defining the unitary rotations, one can optimize the generator for a specific cost function defined on the transformed Hamiltonian. One such cost function is the electronic energy of a product state
\begin{align}
    E(\theta, M; j) =&\bra{\mathrm{HF}} e^{i\theta \hat G}\hat He^{-i\theta \hat G}\ket{\mathrm{HF}}
\end{align}
where $\hat G = \hat V\hat p_j\hat V^\dagger$. 
For each $j$, the matrix $M$ can be optimized classically, using the exact transformation, to determine the generator that yields the largest energy lowering. To minimize the number of terms in the transformed Hamiltonian, $M$ can instead be optimized to reduce the $\ell_0$-norm of the coefficients of the transformed Hamiltonian. In practice, this can be approximated by minimizing the computationally less demanding $\ell_1$-norm.

\section{Conclusions}

We have shown that an exact transformation of many-body operators, such as Hamiltonians, can be achieved when the adjoint action of the generator of the transforming unitary defines a linear map in a finite-dimensional space. According to the Cayley–Hamilton theorem, a finite-degree characteristic polynomial exists that annihilates the adjoint map. This observation enables one to represent the otherwise infinite Baker–Campbell–Hausdorff series for the transformed operator as a finite-degree polynomial, computable with polynomial effort on a classical computer.

We have analyzed two cases in which the adjoint map of the generator can be represented as a linear operator in a finite-dimensional space: (1) when the generator belongs to one of the known finite-dimensional Lie algebras that acts on the space of the transformed operator, and (2) when the generator has a finite number of distinct eigenvalues. In the latter case, the degree of the characteristic polynomial can be reduced by accounting for properties of the transformed operator arising from its commutation and anticommutation relations with the generator. The second case thus subsumes the first and provides a more general framework. Additionally, the second case can also be seen as the generator being the element of a finite abelian Lie algebra of eigen-subspace 
projectors.

When transforming complex operators such as many-body Hamiltonians, one can express them as polynomials in operators whose algebraic relations with the generator simplify the transformation. Particularly efficient transformations arise when both the generator and the operator belong to the same Lie algebra.

Ensuring that the adjoint generator action has finite degree under either of the two conditions is sufficient, as all additional reductions stemming from more specific algebraic structures are recovered automatically through normal ordering of the resulting transformed operators. Consequently, while identifying specific operator fragments is not required to obtain the final result, such analysis explains why, for the same generator, some transformed expressions are more compact than others. For example, in the case of unitary coupled-cluster (UCC) generators, we identify fragments requiring only a single commutator, thereby extending prior results.

The framework introduced here opens several promising directions. A natural extension involves finite-spectrum generators of chemical relevance, such as spherical tensor operators that preserve spin and spatial symmetries. Automated procedures to identify fragments that require fewer nested commutators could further improve the computational efficiency of the exact transformation. Another avenue is to search for new generators that preserve molecular symmetries while having small spectra, thereby providing a useful class of optimizable unitaries for efficient quantum simulation. As an illustration, we construct a class of optimizable, number-conserving generators whose spectra are identical to those of Pauli products.

\section*{Acknowledgments}
The authors thank Smik Patel and Tsung-Chi Chiang for helpful discussions, and Songhao Bao and Ilias Magoulas for constructive comments that helped improve the manuscript. This work was funded under the NSERC Quantum Consortium of Distributed Quantum Computing program and the Applied Quantum Challenge of the National Research Council (NRC).

\section*{Code availablity}
Python scripts to determine the block structures, obtain nested commutator coefficients, and test examples listed in Table \ref{tab:examples} can be found at \href{https://github.com/Praveen91299/UnitaryTransforms.git}{https://github.com/Praveen91299/UnitaryTransforms.git}.

\appendix
\section{Coefficients of nested commutators are real valued}\label{app:reality}
For a Hermitian $\hat H$, $\norm{\hat P_l\hat H\hat P_r}\neq 0$ iff $\norm{\hat P_r\hat H\hat P_l} \neq 0$ and we have $-\Delta \in S$ iff $\Delta \in S$. Then both $\Delta $ and $-\Delta$ define constraint equations through Eq. \eqref{eqn:bch_eqns}. Adding and subtracting these constraints, we obtain an alternative system of equations
\begin{align}
    \frac{e^{i\theta \Delta} + e^{-i\theta\Delta}}{2} =& \sum_{m=0}^{\lfloor (|S| - 1)/2\rfloor} (-1)^mc_{2m}(\theta)\Delta^{2m}\label{eqn:real}\\
    \frac{e^{i\theta\Delta} - e^{-i\theta\Delta}}{2i} =& \sum_{m=0}^{\lfloor|S|/2\rfloor - 1} (-1)^mc_{2m+1}(\theta)\Delta^{2m+1}\label{eqn:imag}
\end{align}
for $\Delta\in \bar S$ from the set of non-negative unique eigenvalue differences
\begin{equation}
    \bar{S} := \{|x| ~|~ x\in S\}.
\end{equation}
Equations \eqref{eqn:real} can be re-expressed as
\begin{align}
    \vec e\,^{(0)} =& W^{(0)}\vec c\,^{(0)}
\end{align}
where $e_\Delta^{(0)} = \cos(\theta \Delta)$, $c_m^{(0)} = c_{2m}(\theta)$ and $[W^{(0)}]_{\Delta m} = (-1)^m\Delta^{2m}$ is the $|\bar S|\times |\bar S|$ dimensional generalized Vandermonde matrix defined by $\Delta \in \bar S$ and $m = 0,\dots, |\bar S|-1$.\cite{gantmacher2025theory} Since Eq. \eqref{eqn:imag} vanishes for $\Delta= 0$, we re-express the equation as
\begin{equation}
    \vec e\,^{(1)} = W^{(1)}\vec c\,^{(1)}
\end{equation}
where $e_\Delta^{(1)} = \sin(\theta \Delta)$, $c_m^{(1)} = c_{2m + 1}(\theta)$, and $[W^{(1)}]_{\Delta m} = (-1)^m\Delta^{2m+1}$ is the $|\bar S \setminus \{0\}|\times |\bar S \setminus \{0\}|$ dimensional generalized Vandermonde matrix defined by $\Delta \in \bar S\setminus\{0\}$ and $m = 0,\dots, |\bar S\setminus \{0\}| - 1$.

The matrices $W^{(0)}, W^{(1)}$ are real, square and invertible, thus yielding unique real solutions for $\vec c\,^{(0)}, \vec c\,^{(1)}$ and equivalently for $\vec c$. Since we have solved a system equations that are equivalent to the original system of equations in Eq. \eqref{eqn:bch_eqns}, the solution must be identical.

\section{Projectors onto the eigenspaces of unitary coupled-cluster generators}\label{app:ucc_proj}

Consider the UCC generator $\hat G$ defined in Eq.~\eqref{eqn:ucc_G}. $\hat G$ has spectrum $\{0,\pm 1\}$, and projectors onto its eigenspaces can be constructed using methods discussed in Ref.~\onlinecite{Izmaylov2019}. Here we present an alternative approach based on enumerating the eigenstates of $\hat G$, which also reveals useful algebraic identities of the involved operators.

Consider a Slater determinant $\ket{\phi_j}$ in a basis of $N$ spin-orbitals. If
\begin{equation}\label{eqn:tg_phi_j}
    \hat T_G \ket{\phi_j} = \ket{\psi_j} \neq 0,
\end{equation}
then $\ket{\psi_j}$ is a Slater determinant distinct from $\ket{\phi_j}$, and the states
\begin{equation}
    \ket{\psi_j^{\pm}} = \frac{1}{\sqrt{2}} \left(\ket{\phi_j} \mp i\,\ket{\psi_j}\right)
\end{equation}
are eigenstates of $\hat G$ with eigenvalues $\pm 1$, respectively. Suppose there are $k$ such pairs $\{\ket{\psi_j^\pm}\}_{j=1}^k$. From Eq.~\eqref{eqn:tg_phi_j},
\begin{equation}
    \sum_{j=1}^k \ket{\psi_j}\bra{\phi_j} = \hat T_G, \label{eqn:tg}
\end{equation}
and consequently,
\begin{align}
    \sum_{j=1}^k \ket{\phi_j}\bra{\psi_j} &= \hat T_G^\dagger, \label{eqn:tg^} \\
    \sum_{j=1}^k \ket{\phi_j}\bra{\phi_j} &= \hat T_G^\dagger \hat T_G, \label{eqn:tg^tg} \\
    \sum_{j=1}^k \ket{\psi_j}\bra{\psi_j} &= \hat T_G \hat T_G^\dagger. \label{eqn:tgtg^}
\end{align}
From Eqs.~\eqref{eqn:tg}–\eqref{eqn:tgtg^}, we have the identities
\begin{align}
    \hat T_G^2 = (\hat T_G^\dagger)^2 &= 0, \label{eqn:tg^2} \\
    \hat T_G \hat T_G^\dagger \hat T_G &= \hat T_G, \label{eqn:tgtg^tg} \\
    \hat T_G^\dagger \hat T_G \hat T_G^\dagger &= \hat T_G^\dagger. \label{eqn:tg^tgtg^}
\end{align}
(These reflect nilpotency and partial-isometry properties of the ladder operators.)

Using Eqs.~\eqref{eqn:tg}–\eqref{eqn:tgtg^}, the projectors onto the $\pm 1$ eigenspaces are
\begin{align}
    \hat P_{\pm} 
    &= \sum_{j=1}^k \ket{\psi_j^\pm}\bra{\psi_j^\pm} \nonumber \\
    &= \frac{1}{2} \sum_{j=1}^k \!\left( \ket{\phi_j}\bra{\phi_j} + \ket{\psi_j}\bra{\psi_j} 
       \pm i\,\ket{\phi_j}\bra{\psi_j} \mp i\,\ket{\psi_j}\bra{\phi_j} \right) \nonumber \\
    &= \frac{1}{2}\!\left( \hat T_G^\dagger \hat T_G + \hat T_G \hat T_G^\dagger \pm i\,\hat T_G^\dagger \mp i\,\hat T_G \right),
\end{align}
and the projector onto the null space is
\begin{equation}
    \hat P_0 = \hat I - \hat P_{+} - \hat P_{-}.
\end{equation}

Define the projector onto the nonzero eigenspaces of $\hat G$ as
\begin{align}
    \hat P_G &:= \hat P_+ + \hat P_- \label{eqn:pg} \\
             &= \hat T_G \hat T_G^\dagger + \hat T_G^\dagger \hat T_G 
              \;=\; \hat G \hat G^\dagger \;=\; \hat G^\dagger \hat G, \label{eqn:pq_equiv}
\end{align}
where the second equality uses $\hat G = (-i)(\hat T_G - \hat T_G^\dagger)$ together with Eq.~\eqref{eqn:tg^2}. The operator $\hat P_G$ projects onto $\mathrm{span}\!\left(\{\ket{\phi_j}, \ket{\psi_j}\}_{j=1}^k\right)$ and is orthogonal to $\hat P_0$ (i.e., $\hat P_G \hat P_0 = 0$). Since the spectrum of $\hat G$ is $\{0,\pm 1\}$, we have the compact relations
\begin{align}
    \hat P_{\pm} &= \frac{1}{2}\left(\hat P_G \pm \hat G\right), \label{eqn:Ppmi} \\
    \hat P_0     &= \hat I - \hat P_G.
\end{align}

\section{Nonzero projected blocks for unitary coupled-cluster generators}\label{app:ucc_blocks}

Consider the case in which $\hat H_\alpha := \hat T_\alpha + \hat T_\alpha^\dagger$ is the sum of a fermionic excitation $\hat T_\alpha = \hat a_C^\dagger \hat a_D$ and its Hermitian conjugate. We determine the set $S$ of unique eigenvalue differences over nonzero projected blocks for all fermionic strings $\hat T_\alpha$. We use the notation introduced in Eqs.~\eqref{eqn:tg_decomp} and \eqref{eqn:ta_decomp}.

First, consider the commuting case $[\hat H_\alpha, \hat G] = 0$. Then $S = \{0\}$. This includes the case in which the excitations $\hat T_\alpha$ and $\hat T_G$ do not share spin-orbital indices ($\tilde T_G = \tilde T_\alpha = \hat I$).

When $\hat T_\alpha$ and $\hat T_G$ share indices, use Eq.~\eqref{eqn:pq_equiv} and consider the block of $\hat H_\alpha$ on the $\pm 1$ eigenspaces:
\begin{align}
    \hat P_G \hat H_\alpha \hat P_G 
    &= \tilde T_G \tilde T_G^\dagger \tilde T_\alpha \tilde T_G \tilde T_G^\dagger \cdot \tilde L_G \tilde L_G^\dagger \tilde L_G \tilde L_G^\dagger \tilde L_\alpha \nonumber\\
    &\quad + \tilde T_G \tilde T_G^\dagger \tilde T_\alpha \tilde T_G^\dagger \tilde T_G \cdot \tilde L_G \tilde L_G^\dagger \tilde L_G^\dagger \tilde L_G \tilde L_\alpha \nonumber\\
    &\quad + \tilde T_G^\dagger \tilde T_G \tilde T_\alpha \tilde T_G \tilde T_G^\dagger \cdot \tilde L_G^\dagger \tilde L_G \tilde L_G \tilde L_G^\dagger \tilde L_\alpha \nonumber\\
    &\quad + \tilde T_G^\dagger \tilde T_G \tilde T_\alpha \tilde T_G^\dagger \tilde T_G \cdot \tilde L_G^\dagger \tilde L_G \tilde L_G^\dagger \tilde L_G \tilde L_\alpha \nonumber\\
    &\quad + \text{h.c.} \label{eqn:pghpg}
\end{align}
(Here, we separate the operator factors over shared and unshared spin-orbitals by “$\cdot$”.) Since
\begin{equation}
    \hat P_G \hat H_\alpha \hat P_G 
    = \hat P_+ \hat H_\alpha \hat P_+ 
    + \hat P_+ \hat H_\alpha \hat P_- 
    + \hat P_- \hat H_\alpha \hat P_+ 
    + \hat P_- \hat H_\alpha \hat P_-,
\end{equation}
if $\hat P_G \hat H_\alpha \hat P_G$ is zero, then all blocks in $\{\hat P_\pm \hat H_\alpha \hat P_\pm, \hat P_\pm \hat H_\alpha \hat P_\mp\}$ vanish. For $\hat P_G \hat H_\alpha \hat P_G$ to be nonzero, at least one product in Eq.~\eqref{eqn:pghpg} must be nonzero, leading to the following cases:
\begin{align}
    \text{(i)}\;& \tilde T_\alpha = \tilde T_G, \quad \tilde L_G = \tilde L_G^{(n)},\\
    \text{(ii)}\;& \tilde T_\alpha = \tilde T_G^\dagger, \quad \tilde L_G = \tilde L_G^{(n)},\\
    \text{(iii)}\;& \tilde T_\alpha = \tilde T_G \tilde T_G^\dagger,\\
    \text{(iv)}\;& \tilde T_\alpha = \tilde T_G^\dagger \tilde T_G,\\
    \text{(v)}\;& \tilde T_\alpha = \tilde T_\alpha^{(n)}, \quad \tilde T_G = \tilde T_G^{(n)}.
\end{align}
Here a superscript $(n)$ denotes a product of occupation operators $\hat n_p := \hat a_p^\dagger \hat a_p$ (including $\hat I$). Case (v) is trivial ($[\hat H_\alpha, \hat G]=0$) and is not considered further. In cases (i)–(iv) we assume $\tilde T_G$ is not a pure number-operator product. In all four cases,
\begin{align}
    [\hat P_G, \hat T_\alpha] 
    &= \tilde T_G \tilde T_G^\dagger \tilde T_\alpha \tilde L_G \tilde L_G^\dagger \tilde L_\alpha 
     - \tilde T_\alpha \tilde T_G \tilde T_G^\dagger \tilde L_G \tilde L_G^\dagger \tilde L_\alpha \nonumber\\
    &\quad + \tilde T_G^\dagger \tilde T_G \tilde T_\alpha \tilde L_G^\dagger \tilde L_G \tilde L_\alpha 
     - \tilde T_\alpha \tilde T_G^\dagger \tilde T_G \tilde L_G^\dagger \tilde L_G \tilde L_\alpha \;=\; 0,\label{eq:ucc_exc_comm}
\end{align}
so $[\hat P_G,\hat H_\alpha]=0$ and therefore
\begin{align}
    \hat P_G \hat H_\alpha \hat P_0 &= \hat H_\alpha \hat P_G \hat P_0 = 0, \\
    \Rightarrow\quad \hat P_\pm \hat H_\alpha \hat P_0 &= \hat P_0 \hat H_\alpha \hat P_\pm = 0.
\end{align}
Thus either the set $\{\hat P_\pm \hat H_\alpha \hat P_\pm, \hat P_\pm \hat H_\alpha \hat P_\mp\}$ or the set $\{\hat P_\pm \hat H_\alpha \hat P_0, \hat P_0 \hat H_\alpha \hat P_\pm\}$ can be nonzero, but not both simultaneously. Consequently, $S = \{0,\pm 2\}$ or $S=\{0,\pm 1\}$, so that $|S|-1 = 2$ nested commutators suffice to represent $\hat H_\alpha'$.

Note that cases (i)-(iv) also correspond to cases when $\hat P_G\hat T_\alpha\hat P_G$ is nonzero. From Eq.~\eqref{eq:ucc_exc_comm}, it follows that either the set $\{\hat P_\pm \hat T_\alpha \hat P_\pm, \hat P_\pm \hat T_\alpha \hat P_\mp\}$ or the set $\{\hat P_\pm \hat T_\alpha \hat P_0, \hat P_0 \hat T_\alpha \hat P_\pm\}$ can be nonzero, but not both simultaneously. Consequently, $S = \{0,\pm 2\}$ or $S=\{0,\pm 1\}$ for a single Fermionic excitation.

\medskip
\noindent\textbf{When $\mathbf{\hat P_G \hat H_\alpha \hat P_G \neq 0}$.}
Using the partial-isometry identities $\tilde T_G \tilde T_G^\dagger \tilde T_G = \tilde T_G$ and $\tilde T_G^\dagger \tilde T_G \tilde T_G^\dagger = \tilde T_G^\dagger$ in cases (i) and (ii), we have
\begin{align}
    \hat P_0 \hat H_\alpha \hat P_0 
    &= \hat H_\alpha ( \hat I - \hat P_G ) \nonumber\\
    &= \hat H_\alpha \big(\hat I - \tilde L_G^{(n)}\big)
\end{align}
Whereas in case (iii) we have
\begin{align}
    \hat P_0 \hat H_\alpha \hat P_0 &= \hat H_\alpha \big(\hat I - \tilde L_G \tilde L_G^\dagger\big),
\end{align}
and in case (iv) we have
\begin{align}
    \hat P_0 \hat H_\alpha \hat P_0 &= \hat H_\alpha \big(\hat I - \tilde L_G^\dagger \tilde L_G\big).
\end{align}
In all four cases, when $\tilde L_G = \hat I$, $\hat P_0 \hat H_\alpha \hat P_0$ vanishes; otherwise it is nonzero.

The operator $\tilde L_\alpha$ remains unspecified; its form may allow further reductions in the number of nonzero projected blocks. If the off-diagonal blocks $\{\hat P_\pm \hat H_\alpha \hat P_\mp\}$ vanish, then $[\hat H_\alpha,\hat G]=0$, which we do not consider further here. To test whether the diagonal blocks $\{\hat P_\pm \hat H_\alpha \hat P_\pm\}$ can vanish, we examine the anti-commutator $[\hat H_\alpha,\hat G]_+$ for cases (i)–(iv).

Case (i): $\tilde T_\alpha = \tilde T_G$ and $\tilde L_G = \tilde L_G^{(n)}$. $\tilde T_G$ is a product of an even number of fermionic operators (since $\tilde L_G = \tilde L_G^{(n)}$), and the operators $\tilde T_G, \tilde L_G^{(n)}$ and $\tilde L_\alpha$ do not share spin-orbital indices, so
\begin{equation}
    [\tilde T_G, \tilde L_\alpha] = [\tilde T_G, \tilde L_G^{(n)}] = [\tilde L_G^{(n)}, \tilde L_\alpha] = 0.
\end{equation}
Then
\begin{align}
    \hat H_\alpha \hat G 
    &= i \big( \tilde T_G \tilde T_G^\dagger \tilde L_G^{(n)} \tilde L_\alpha 
             - \tilde T_G^\dagger \tilde T_G \tilde L_G^{(n)} \tilde L_\alpha^\dagger \big), \\
    \Rightarrow\quad [\hat H_\alpha, \hat G]_+ 
    &= i \big( \tilde T_G^\dagger \tilde T_G + \tilde T_G \tilde T_G^\dagger \big)\, \tilde L_G^{(n)} \big( \tilde L_\alpha - \tilde L_\alpha^\dagger \big). \label{eqn:case_i_anti}
\end{align}
When $\tilde L_\alpha = \tilde L_\alpha^{(n)}$ is a number-operator product, $\tilde L_\alpha^\dagger = \tilde L_\alpha$ and the anti-commutator in Eq.~\eqref{eqn:case_i_anti} vanishes, implying the diagonal blocks $\{\hat P_\pm \hat H_\alpha \hat P_\pm\}$ vanish. Moreover, when $\tilde L_G = \hat I$, $\hat P_0 \hat H_\alpha \hat P_0$ vanishes and $S=\{\pm 2\}$, so $|S|=2$.

Case (ii): $\tilde T_\alpha = \tilde T_G^\dagger$ and $\tilde L_G = \tilde L_G^{(n)}$. Then
\begin{align}
    \hat H_\alpha \hat G 
    &= i \big( \tilde T_G^\dagger \tilde T_G \tilde L_G^{(n)} \tilde L_\alpha^\dagger 
             - \tilde T_G \tilde T_G^\dagger \tilde L_G^{(n)} \tilde L_\alpha \big), \\
    \Rightarrow\quad [\hat H_\alpha, \hat G]_+ 
    &= i \big( \tilde T_G^\dagger \tilde T_G + \tilde T_G \tilde T_G^\dagger \big)\, \tilde L_G^{(n)} \big( \tilde L_\alpha^\dagger - \tilde L_\alpha \big),
\end{align}
which is analogous to case (i), yielding $S=\{\pm 2\}$ for $\tilde L_\alpha = \tilde L_\alpha^{(n)}$ and $\tilde L_G^{(n)} = \hat I$.

Case (iii): $\tilde T_\alpha = \tilde T_G \tilde T_G^\dagger$. Then
\begin{align}
    \hat H_\alpha \hat G 
    &= -i \big( \tilde L_\alpha + \tilde L_\alpha^\dagger \big)\, \tilde T_G \tilde L_G, \\
    \Rightarrow\quad [\hat H_\alpha, \hat G]_+ 
    &= -i \big( \tilde L_\alpha + \tilde L_\alpha^\dagger \big)\, \big( \tilde T_G \tilde L_G - \tilde L_G^\dagger \tilde T_G^\dagger \big).
\end{align}
Since $[\hat H_\alpha,\hat G]_+$ cannot vanish for any choice of $\tilde L_\alpha$ in this case, at least one of the diagonal blocks $\{\hat P_\pm \hat H_\alpha \hat P_\pm\}$ is nonzero, and no further reduction of $S$ is possible.

Case (iv): $\tilde T_\alpha = \tilde T_G^\dagger \tilde T_G$. Then
\begin{align}
    \hat H_\alpha \hat G 
    &= i \big( \tilde L_\alpha + \tilde L_\alpha^\dagger \big)\, \tilde L_G^\dagger \tilde T_G^\dagger, \\
    \Rightarrow\quad [\hat H_\alpha, \hat G]_+ 
    &= -i \big( \tilde L_\alpha + \tilde L_\alpha^\dagger \big)\, \big( \tilde T_G \tilde L_G - \tilde L_G^\dagger \tilde T_G^\dagger \big),
\end{align}
which is analogous to case (iii) and likewise does not permit further reduction of $S$.

\medskip
\noindent\textbf{When $\mathbf{\hat P_G \hat H_\alpha \hat P_G = 0}$.}
The blocks $\{\hat P_0 \hat H_\alpha \hat P_0, \hat P_\pm \hat H_\alpha \hat P_0, \hat P_0 \hat H_\alpha \hat P_\pm\}$ can potentially be nonzero. Since $\hat H_\alpha$ is Hermitian, $\hat P_\pm \hat H_\alpha \hat P_0$ is nonzero iff $\hat P_0 \hat H_\alpha \hat P_\pm$ is nonzero. Hence the only avenues for reducing $S$ are: (i) all blocks in $\{\hat P_\pm \hat H_\alpha \hat P_0, \hat P_0 \hat H_\alpha \hat P_\pm\}$ vanish (the commuting case $[\hat H_\alpha,\hat G]=0$, not considered further), or (ii) $\hat P_0 \hat H_\alpha \hat P_0$ vanishes. For the latter, we compute
\begin{align}
    \hat P_0 \hat H_\alpha \hat P_0 
    &= \tilde T_\alpha \tilde L_\alpha \nonumber\\
    &\quad - \tilde T_G^\dagger \tilde T_G \tilde T_\alpha \cdot \tilde L_G^\dagger \tilde L_G \tilde L_\alpha \label{eqn:doub_1}\\
    &\quad - \tilde T_G \tilde T_G^\dagger \tilde T_\alpha \cdot \tilde L_G \tilde L_G^\dagger \tilde L_\alpha \label{eqn:doub_2}\\
    &\quad - \tilde L_G^\dagger \tilde L_G \tilde L_\alpha^\dagger \cdot \tilde T_G^\dagger \tilde T_G \tilde T_\alpha^\dagger \label{eqn:doub_3}\\
    &\quad - \tilde L_G \tilde L_G^\dagger \tilde L_\alpha^\dagger \cdot \tilde T_G \tilde T_G^\dagger \tilde T_\alpha^\dagger \label{eqn:doub_4}\\
    &\quad + \text{h.c.}
\end{align}
(Here, the first line collects the direct term; “$\cdot$” separates operator factors over shared and unshared spin-orbital indices.) If only one of the terms in Eqs.~\eqref{eqn:doub_1}–\eqref{eqn:doub_4} is nonzero, then $\hat P_0 \hat H_\alpha \hat P_0$ cannot vanish. For example, if Eq.~\eqref{eqn:doub_1} is nonzero,
\begin{equation}
    \hat P_0 \hat H_\alpha \hat P_0 = \big( \hat I - \hat T_G^\dagger \hat T_G \big)\, \hat T_\alpha + \text{h.c.}
\end{equation}
is nonzero. The same reasoning applies if only Eq.~\eqref{eqn:doub_2}, \eqref{eqn:doub_3}, or \eqref{eqn:doub_4} is nonzero.

Next, consider nonzero \emph{pairs} among Eqs.~\eqref{eqn:doub_1}–\eqref{eqn:doub_4}. Nonzero \eqref{eqn:doub_1} and \eqref{eqn:doub_3} require $\tilde T_\alpha = \tilde T_G^\dagger \tilde T_G$ (case (iv)), which implies $\hat P_G \hat H_\alpha \hat P_G \neq 0$. Similarly, nonzero \eqref{eqn:doub_2} and \eqref{eqn:doub_4} require $\tilde T_\alpha = \tilde T_G \tilde T_G^\dagger$ (case (iii)). There is no nontrivial $\tilde T_\alpha$ for which both \eqref{eqn:doub_1} and \eqref{eqn:doub_2} are nonzero, and likewise \eqref{eqn:doub_3} and \eqref{eqn:doub_4} cannot both be nonzero. 

When $\tilde T_\alpha = \tilde T_G$, Eqs.~\eqref{eqn:doub_2} and \eqref{eqn:doub_3} can be nonzero, and
\begin{align}
    \hat P_0 \hat H_\alpha \hat P_0 
    &= \big( \tilde T_G \tilde L_\alpha + \tilde L_\alpha^\dagger \tilde T_G^\dagger \big) 
       \big( \hat I - \tilde L_G^\dagger \tilde L_G - \tilde L_G \tilde L_G^\dagger \big).
\end{align}
If $\tilde L_G^\dagger \tilde L_G + \tilde L_G \tilde L_G^\dagger = \hat I$, then $\hat P_0 \hat H_\alpha \hat P_0 = 0$. This occurs when $\tilde L_G \in \{\hat a_j^\dagger, \hat a_j\}$ is a single creation or annihilation operator on some spin-orbital index $j \notin C \cup D$. Similarly, when $\tilde T_\alpha = \tilde T_G^\dagger$, Eqs.~\eqref{eqn:doub_1} and \eqref{eqn:doub_4} can be nonzero and
\begin{align}
    \hat P_0 \hat H_\alpha \hat P_0 
    &= \big( \tilde L_\alpha^\dagger \tilde T_G + \tilde T_G^\dagger \tilde L_\alpha \big) 
       \big( \hat I - \tilde L_G^\dagger \tilde L_G - \tilde L_G \tilde L_G^\dagger \big),
\end{align}
which again vanishes when $\tilde L_G$ is a single creation or annihilation operator.

Because the nonzero pairs among Eqs.~\eqref{eqn:doub_1}–\eqref{eqn:doub_4} are disjoint, there is no single choice of $\hat T_\alpha,\hat T_G$ for which three products are nonzero; having all four nonzero would require $\tilde T_\alpha = \tilde T_G = \tilde T_G^\dagger$, i.e., the trivial case (v).

\section{Disjoint classes of fermionic involutory generators}\label{app:ferm_refl}

Here, we show that the fermionic operators $\{\hat p_{j}\}_{j=1}^{10}$ in Eq.~\eqref{eq:refl_poly_all}, defined as polynomials in $\{\hat n_p\}$, cannot be transformed into one another by orbital-rotation unitaries [Eq.~\eqref{eqn:orbital_rotation}]. We first identify properties of these polynomials that remain invariant under orbital rotations. Polynomials with distinct values of these invariants cannot be related by orbital rotations. For the remaining cases with identical invariant values, we explicitly show that no orbital rotation can transform one polynomial into another.

We begin by noting that orbital rotations can be chosen to permute the indices defining a given polynomial. This allows us to group together polynomials that are related by index relabelling.
\begin{lemma}
    There always exists an orbital rotation $\hat V$ whose action on $p(\{n_p\})$ yields $p(\{n_{P(p)}\})$ for some permutation $P$ of the indices.\label{lemma:permutation}
\end{lemma}
\begin{proof}
    The orbital rotation
    \begin{equation}
        \hat V = \exp\left[i\frac{\pi}{2}\bigl(\hat a_p^\dagger \hat a_q + \hat a_q^\dagger \hat a_p\bigr)\right]
    \end{equation}
    swaps $\hat n_p$ and $\hat n_q$ while leaving $\hat n_r$ for $r\neq p,q$ unchanged. Since any permutation can be expressed as a product of swaps, we can always construct an orbital rotation that implements the desired permutation.
\end{proof}

The fermionic rank $r(\hat O)$ of a normal-ordered fermionic string $\hat O$ consisting of an equal number of creation and annihilation operators is defined as the number of creation–annihilation pairs. When $\hat O$ is a sum of fermionic strings, we define its rank to be the maximum of the ranks of the individual strings. For instance, both $\hat a_i^\dagger \hat a_j^\dagger \hat a_k \hat a_l$ and $\hat a_i^\dagger \hat a_j^\dagger \hat a_k \hat a_l + \hat a_i^\dagger \hat a_j$ have rank $2$.
\begin{lemma}
    The fermionic rank of normal-ordered operators is preserved under orbital rotations. Equivalently, if two normal-ordered fermionic operators $\hat A$ and $\hat B$ have different fermionic ranks, then no orbital rotation $\hat V$ satisfies $\hat V\hat A\hat V^\dagger = \hat B$.
    \label{lemma:ferm_rank}
\end{lemma}
\begin{proof}
    Let $\hat A = \hat a_{i_1}^\dagger \cdots \hat a_{i_N}^\dagger \hat a_{j_1}\cdots \hat a_{j_N}$ be a fermionic string of rank $N$. The transformed operator is of the form
    \begin{equation}
        \hat V\hat A\hat V^\dagger = \sum_{\mathclap{p_1,\ldots,p_N,\, q_1,\ldots,q_N}}
        V_{p_1 i_1}\cdots V_{p_N i_N}
        V_{q_1 j_1}^* \cdots V_{q_N j_N}^*
        \hat a_{p_1}^\dagger \cdots \hat a_{p_N}^\dagger \hat a_{q_1}\cdots \hat a_{q_N},
    \end{equation}
    where $V = \exp[iM]$ is the single-particle representation of the orbital rotation $\hat V$. Terms with $p_i=p_j$ or $q_i=q_j$ for some $i\neq j$ vanish, since $(\hat a_{p_i})^2 = (\hat a_{p_i}^\dagger)^2 = 0$. The surviving terms all contain distinct indices, $p_i\neq p_j$ and $q_i\neq q_j$ for all $i\neq j$. Normal ordering these terms only introduces an overall sign in $\{\pm 1\}$ and leaves the number of creation–annihilation pairs equal to $N$. There cannot be a reduction in rank under the transformation, since creation operators always precede annihilation operators in the normal-ordered fermionic strings. These arguments extend directly to operators that are sums of fermionic strings, by restricting attention to the highest-rank contributions in $\hat A$ and $\hat B$.
\end{proof}

Rank equivalence is therefore a necessary condition for two operators to be related by an orbital rotation. Consequently, if an orbital rotation acts on a linear combination of normal-ordered fermionic operators with multiple ranks, it must map the contribution at each rank to a contribution of the same rank.
\begin{lemma}
    Let $\hat A = \sum_k \hat A_k$ and $\hat B = \sum_k \hat B_k$ be sums of normal-ordered fermionic operators with $r(\hat A_k) = r(\hat B_k) = k$. If an orbital rotation $\hat V$ satisfies $\hat V\hat A\hat V^\dagger = \hat B$, then $\hat V\hat A_k\hat V^\dagger = \hat B_k$ for all $k$, and $\mathrm{Tr}(\hat A_k) = \mathrm{Tr}(\hat B_k)$.
    \label{lemma:sub_ferm_rank}
\end{lemma}
\begin{proof}
    By Lemma~\ref{lemma:ferm_rank}, $\hat V\hat A_k\hat V^\dagger$ has rank $k$. Since operators of different rank cannot coincide, the result follows.
\end{proof}

Normal ordering of arbitrary fermionic strings may produce sums of strings of multiple ranks. In contrast, products of $\{\hat n_p\}$ remain single strings under normal ordering. This allows us to extend Lemma~\ref{lemma:sub_ferm_rank} to polynomials in $\{\hat n_p\}$.
\begin{lemma}\label{lemma:rank_poly}
    Fermionic rank is preserved under orbital rotations of homogeneous polynomials in $\{\hat n_p\}$. Furthermore, suppose $\hat V\hat A\hat V^\dagger = \hat B$ for $\hat A = \sum_k \hat A_k$ and $\hat B = \sum_k \hat B_k$, where $\hat A_k$ and $\hat B_k$ are homogeneous polynomials in $\{\hat n_p\}$ of rank $k$. Then $\hat V\hat A_k\hat V^\dagger = \hat B_k$ for all $k$, and $\mathrm{Tr}(\hat A_k) = \mathrm{Tr}(\hat B_k)$.
\end{lemma}
\begin{proof}
    Each $\hat n_p = \hat a_p^\dagger \hat a_p$ is already normal-ordered, and products of $\{\hat n_p\}$ contain no repeated indices in our construction. Normal ordering such a product yields a single fermionic string of fixed length. By Lemma~\ref{lemma:ferm_rank}, orbital rotations preserve the fermionic rank of this string, and the statement follows.
\end{proof}

Thus, the trace of a polynomial in $\{\hat n_p\}$, as well as the traces of its individual homogeneous components, are invariant under orbital rotations. The ten reflection polynomials $\hat p_i$ can therefore be separated based on their traces and on the traces of their homogeneous components (listed in Table~\ref{tab:trace}). To define the trace, we must specify the underlying Fock space. By Lemma~\ref{lemma:permutation}, it is sufficient to work in a four-mode Fock space with indices $(p, q, r, s) = (0, 1, 2, 3)$.

\begin{table}[]
    \centering
    \begin{tabular}{|c|c|c|c|}
    \hline
    Class & Fermionic rank(s) & Trace & Fixed-rank trace \\ \hline
    $\hat p_{1}$ & $\{0, 1\}$ & 0 & (16, -16, 0) \\
    $\hat p_{2}$ & $\{0, 2\}$ & 8 & (16, 0, -8) \\
    $\hat p_{3}$ & $\{0, 1, 2\}$ & 8 & (16, -16, 8) \\
    $\hat p_{4}$ & $\{0, 1, 2\}$ & -8 & (16, -32, 8) \\
    $\hat p_{5}$ & $\{0, 1, 2\}$ & 0 & (16, -32, 16) \\
    $\hat p_{6}$ & $\{0, 1, 2\}$ & 0 & (16, -16, 0) \\
    $\hat p_{7}$ & $\{0, 1, 2\}$ & 8 & (16, -16, 8) \\
    $\hat p_{8}$ & $\{0, 1, 2\}$ & 0 & (16, -32, 16) \\
    $\hat p_{9}$ & $\{0, 1, 2\}$ & -8 & (16, -48, 24) \\
    $\hat p_{10}$ & $\{0, 1, 2\}$ & 0 & (16, -32, 16)\\\hline
\end{tabular}
    \caption{Traces of the quadratic fermionic reflection polynomials over the four-mode Fock space $\mathcal F = \bigoplus_{j=0}^4 \mathcal F_j$. The column ``Fixed-rank trace'' lists the traces of the homogeneous components of ranks $0$, $1$, and $2$.}
    \label{tab:trace}
\end{table}

It is immediate that $\hat p_{1}$ and $\hat p_{2}$ cannot be converted into any other class, because they lack homogeneous components of both ranks $1$ and $2$. From the traces, it remains to explicitly show that the following classes are disjoint:
\begin{enumerate}
    \item $\hat p_{3}$ and $\hat p_{7}$,
    \item $\hat p_{5}$, $\hat p_{8}$, and $\hat p_{10}$.
\end{enumerate}

\textbf{Case 1: $\hat p_{3}$ and $\hat p_{7}$.}\\
Here $\hat p_{3} = 1 - 2\hat n_p +  2\hat n_p\hat n_q$ and $\hat p_{7} = 1 - 2\hat n_p - 2\hat n_q\hat n_r + 2\hat n_p\hat n_r + 2\hat n_p\hat n_q$. If $\hat V\hat p_3\hat V^\dagger = \hat p_7$, then by Lemma~\ref{lemma:rank_poly},
\begin{equation}
    \hat V \hat n_p\hat n_q\hat V^\dagger = - \hat n_q\hat n_r + \hat n_p\hat n_r + \hat n_p\hat n_q.
\end{equation}
The left-hand side is a projector because $\hat n_p\hat n_q$ is idempotent. The right-hand side is not idempotent, so no such orbital rotation $\hat V$ can exist.

\textbf{Case 2: $\hat p_{5}$, $\hat p_{8}$, and $\hat p_{10}$.}\\[2pt]
\noindent\textbf{Case 2(a): $\hat p_{5}$ and $\hat p_{8}$.}\\
The same type of argument as in Case~1 applies. Assuming $\hat V \hat p_{5}\hat V^\dagger = \hat p_{8}$ and using Lemma~\ref{lemma:rank_poly} leads to a rank-2 projector on the left-hand side being mapped to a non-idempotent operator on the right-hand side, which is impossible.

\noindent\textbf{Case 2(b): $\hat p_{5}$ and $\hat p_{10}$.}\\
Again, the same argument as in Case~1 shows that an orbital rotation mapping $\hat p_{5}$ to $\hat p_{10}$ cannot exist.

\noindent\textbf{Case 2(c): $\hat p_{8}$ and $\hat p_{10}$.}\\
Suppose there exists an orbital rotation $\hat V$ such that $\hat V \hat p_{8}\hat V^\dagger = \hat p_{10}$. Then, by Lemma~\ref{lemma:rank_poly}, the rank-2 parts must satisfy
\begin{equation}\label{eqn:rank2_p27_p29}
    \hat V \bigl(\hat n_p\hat n_r + \hat n_p\hat n_q\bigr)\hat V^\dagger = -\hat n_r\hat n_s +\hat n_p\hat n_r + \hat n_q\hat n_s + \hat n_p\hat n_q.
\end{equation}
Squaring Eq.~\eqref{eqn:rank2_p27_p29}, we obtain
\begin{align}
    &\hat V\bigl(\hat n_p\hat n_r + \hat n_p\hat n_q + 2\hat n_p\hat n_q\hat n_r\bigr)\hat V^\dagger =\nonumber \\ 
    &\hat n_r\hat n_s +\hat n_p\hat n_r + \hat n_q\hat n_s + \hat n_p\hat n_q - 2\hat n_p\hat n_r\hat n_s - 2\hat n_q\hat n_r\hat n_s\nonumber \\ 
    &\quad + 2\hat n_p\hat n_q\hat n_r + 2\hat n_p\hat n_q\hat n_s.\label{eqn:square_p27_p29}
\end{align}
By Lemma~\ref{lemma:rank_poly}, the rank-2 parts of Eq.~\eqref{eqn:square_p27_p29} must individually transform into one another,
\begin{align}
    \hat V(\hat n_p\hat n_r + \hat n_p\hat n_q)\hat V^\dagger =& \hat n_r\hat n_s +\hat n_p\hat n_r + \hat n_q\hat n_s + \hat n_p\hat n_q.
\end{align}
Since $\hat V$ is unitary, we require
\begin{align}
    \mathrm{Tr}(\hat n_p\hat n_r + \hat n_p\hat n_q)
    &= \mathrm{Tr}(\hat n_r\hat n_s +\hat n_p\hat n_r + \hat n_q\hat n_s + \hat n_p\hat n_q) \\
    &\Rightarrow \mathrm{Tr}(\hat n_r\hat n_s + \hat n_q\hat n_s) = 0,
\end{align}
which is false for distinct indices $p, q, r, s$. Therefore, no such orbital rotation exists.


\begin{thebibliography}{32}
\bibitem [{not()}]{note:BCH}%
  \BibitemOpen
  \href@noop {} {}\bibinfo {note} {Formally, the series used here corresponds
  to the Campbell identity, which is a special case of the more general
  Baker--Campbell--Hausdorff relation. We retain the conventional term `BCH
  expansion,' as it is standard in physics and quantum chemistry.}\BibitemShut
  {Stop}%
\end{thebibliography}
%

\end{document}